\documentclass[preprintnumbers,amsmath,amssymb,twocolumn]{revtex4-1}
\usepackage{graphicx}
\usepackage{dcolumn}
\usepackage{bm}
\usepackage{physics}
\usepackage[caption=false]{subfig}

\begin{document}

\title{Spin-mechanics with levitating ferromagnetic particles}

\author{P. Huillery$^{1}$}
\author{T. Delord$^{1}$\footnote{Both authors contributed equally to this work}}
\author{L. Nicolas$^{1}$}
\author{M. Van Den Bossche$^{1}$}
\author{M. Perdriat$^{1}$}
\author{G. H\'etet$^{1}$} 

\affiliation{$^1$Laboratoire de Physique de l'Ecole normale sup\'erieure, ENS, Universit\'e PSL, CNRS, Sorbonne Universit\'e, Universit\'e Paris-Diderot, Sorbonne Paris Cit\'e, Paris, France.}

\begin{abstract}
We propose and demonstrate first steps towards schemes where the librational mode of levitating ferromagnets can be strongly coupled to the electronic spin of Nitrogen-Vacancy (NV) centers in diamond.
Experimentally, we levitate soft ferromagnets in a Paul trap and employ magnetic fields to attain oscillation frequencies in the hundreds of kHz range with Q factors close to $10^4$. These librational frequencies largely exceed the decoherence rate of NV centers in typical CVD grown diamonds offering prospects for  resolved sideband operation. We also prepare and levitate composite diamond-ferromagnet particles and demonstrate both coherent spin control of the NV centers and read-out of the particle libration using the NV spin. Our results will find applications in ultra-sensitive gyroscopy and bring levitating objects a step closer to spin-mechanical experiments at the quantum level.
\end{abstract}

\maketitle

Spin-mechanical systems where electronic spins are entangled to the motion of individual atoms are now widely used for studying fundamental phenomena and for quantum information and metrological applications \cite{Leibfried}. 
Inspired by the coupling schemes developed for trapped atoms \cite{PhysRevLett.67.181} and single ions \cite{Monroe1131,Myatt2000DecoherenceOQ}, new ideas emerged to extend the field to macroscopic systems \cite{PhysRevLett.100.136802,Scala, Wan,chen19, PhysRevLett.99.140403, PhysRevB.79.041302, 2013OExpr..2129695Z, Yin2015, PhysRevLett.117.015502, delord2017strong, 2017NatCo...814358M,Abdi17}, with perspectives to test quantum mechanics on a large scale \cite{Bassi2013}. 

One long-standing goal in the field is the coupling of a macroscopic object to electronic spins. 
In this direction, the electronic spin of the nitrogen-vacancy (NV) center in diamond stands out as a promising solid state qu-bit with efficient optical initialization and read-out \cite{Gruber} and long coherence times \cite{balasubramanian2009ultralong, Maurer1283, BarGill}.
Proposals for coupling NV spins to the motion of cantilevers at the quantum level through magnetic forces \cite{PhysRevB.79.041302,Rabl2,Bennett} or lattice strain \cite{PhysRevB.88.064105,2017NatCo...814358M} have been put forward.
Although important experimental achievements have been made \cite{Arcizet, Kolkowitz, PhysRevLett.111.227602, 2014PhRvL.113b0503T, Ovartchaiyapong2014, MacQuarrie15, Barfuss2015, 2016PhRvL.116n3602G}, the low spin-mechanical coupling rate caused by the large mass of the cantilever has for now not allowed coherent actuation and cooling of its motion.
Levitating nano- or micro-diamonds provide light mechanical oscillators and alternative schemes have been proposed to couple their center of mass \cite{PhysRevA.88.033614,Scala} or librations \cite{Ma,delord2017strong} to NV spins. Under high vacuum, such platforms could take advantage of the high quality factor of the mechanical oscillator \cite{Chang19012010, Romero2}, but experimental progresses with optical tweezers \cite{Neukirch, Hoang, PhysRevLett.117.123604} 
have been hampered by laser heating of the diamond \cite{Rahman,Neukirch,Hoang}. Other scattering-free traps such as magneto-gravitational \cite{Hsu} or Paul \cite{Kuhlicke,delord2016,PhysRevLett.121.053602,Conangla} traps have also been implemented. However, their low mechanical frequency put a severe road-block on the way towards recently proposed spin-entanglement schemes \cite{chen19,Rabl3} and limits the efficiency of the recently demonstrated spin-cooling of the motion of a trapped diamond \cite{DelordNature}.

\begin{figure}
  \centering \scalebox{0.34}{\includegraphics{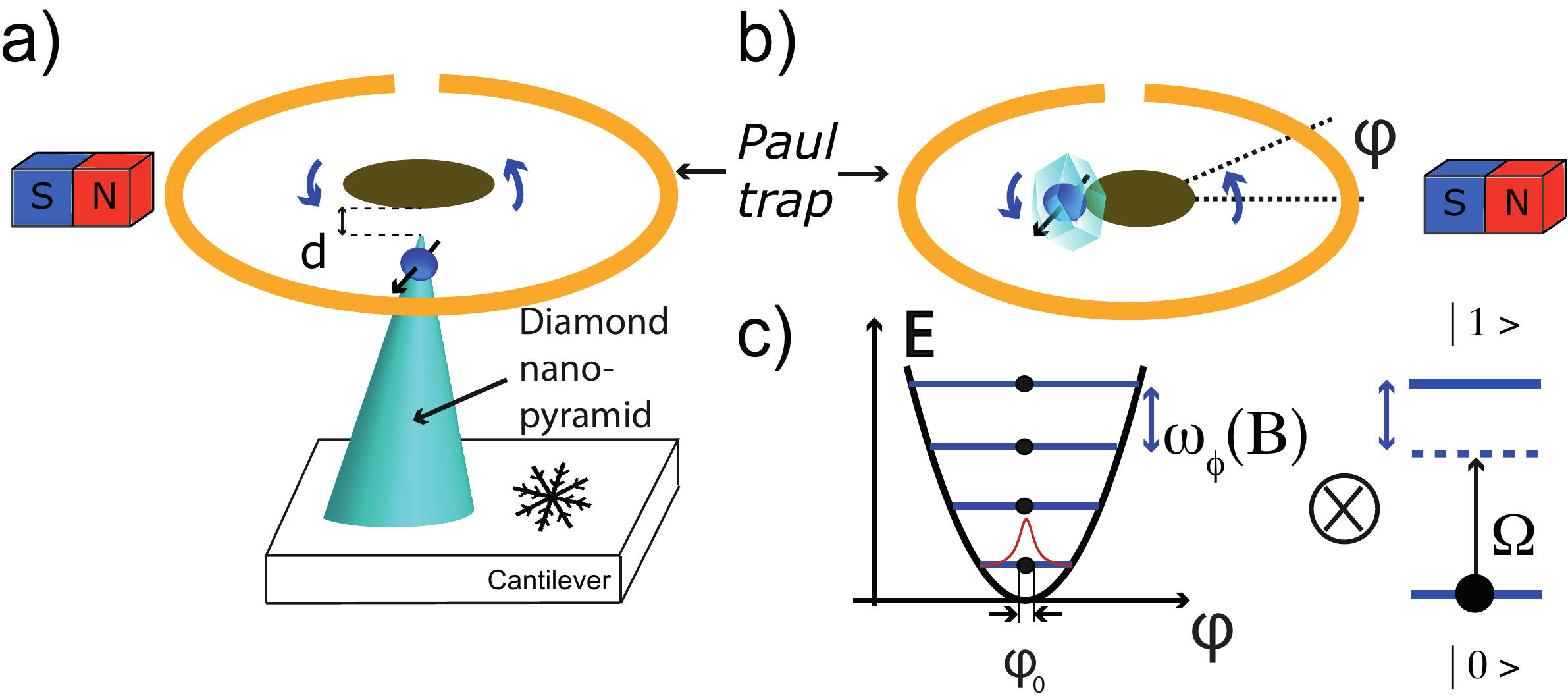}}
  \caption{Schematics showing two proposed platforms for strong spin-mechanical coupling using the librational motion of ferromagnets levitating in a Paul trap. In the first setup (a), a nano-magnet is levitating at a distance $d$ from the electronic spin of an NV center within a bulk diamond at cryogenic temperature. 
In the second proposed set-up (b) a diamond is attached to a nano-magnet and coupled to the libration {\it via} the external static $B$ field.  
  c) Potential energy of the magnetic field dependent harmonic libration resonantly coupled to the two-level spin system via a detuned microwave.
 }\label{plat}
\end{figure}

Here we propose and demonstrate first steps towards novel platforms for spin-mechanical coupling that leverages the aforementioned issues. Here, ferromagnets are levitated in a Paul trap and their librational mode is confined by an external magnetic field. Under modest magnetic fields (in the 0.1 to 1 T range), this levitated oscillator can reach frequencies ranging from 0.1 to 1 MHz, similar to the mechanical frequencies obtained using optical tweezers. 

The proposed spin-mechanical coupling is achieved in the two ways that are depicted in Fig. \ref{plat}. 
In the first scheme, the libration of a ferromagnet is coupled to the spin of a distant ($\sim 1 \mu m$) NV center located in a fixed cold nanopyramid (Fig. 1-a)) grown by Chemical Vapor Deposition (CVD), inspired by magnetic force microscopy (MRFM) \cite{Degen1313, Rugar2004} and single spin magnetometry platforms \cite{rondin, Kolkowitz}. This scheme allows harnessing of the properties of NV spins in pristine diamonds at cryogenic temperature, in particular high fidelity initialization \cite{robledo2011high}, and high fidelity projective read-out through heralding protocols \cite{rao2016heralded,Bennett}. One downside of this proposal is the necessity to control the position of a diamond tip at the nanometer level and 
the potential nuisance coming from the interaction between the charges on the two objects which may impact the magnet motion.  
In the second scheme, a diamond containing NV centers is attached directly to the levitating ferromagnet (Fig.~1-b)). This scheme will bypass the above mentioned distant coupling constraints. Furthermore, although it is similar to the proposal put forward in \cite{delord2017strong} it will benefit from the large librational frequencies of the hybrid structure, which is necessary for resolved sideband operations. 

Towards the first scheme, we assemble and levitate micron-sized particles of soft ferromagnetic materials experimentally. Under a magnetic field of 0.1 T, we observe librational frequencies exceeding 150~kHz, close to the decoherence rate of NV centers in pure diamonds, and $Q$-factors close to $10^4$ at only $10^{-2}$ mbar of vacuum pressure. 
Towards the second scheme, we trap two types of hybrid particles composed of a nano-diamond attached to a soft ferromagnet and a micro-diamond with a ferromagnetic coating. Using these composite structures with large librational confinement, we show both efficient coherent manipulations of the NV centers's spins and spin read-out of the particle libration.

\section{The levitating magnets}
\begin{figure}[ht!]
\centerline{\scalebox{0.24}{\includegraphics{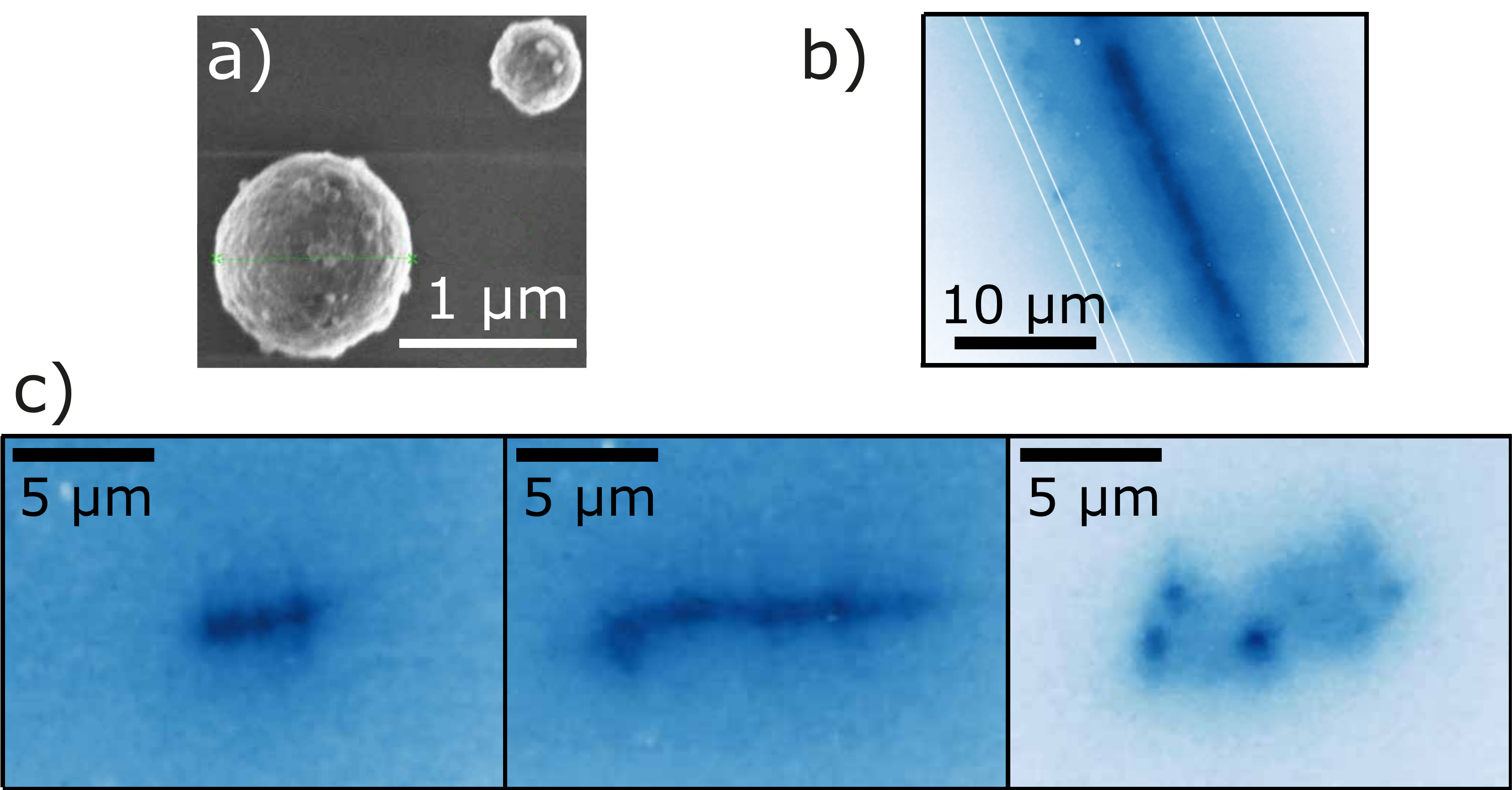}}}
\caption{a) Scanning electron microscopy image of the iron particles outside the trap. b) Image of the 15 $\mu$m diameter wire used to determine the magnification of the imaging system.  c) Images of 3 levitating rods of various shapes.}
\label{rods-images}
\end{figure}

At the heart of these proposals, is the levitation of ferromagnetic particles in a Paul trap. 
In this section, we first demonstrate the levitation of spherical micron-sized particles made from iron with 98~\% purity (Goodfellow, ref. FE006045), using similar ring traps and injection techniques than in \cite{vacuumESR,PhysRevLett.121.053602}. 
The trap consists in a small 25~$\mu$m thick tungsten wire with an inner radius of 200~$\mu$m. It is oriented so that the ring plane is perpendicular to the optical axis. As shown in the SEM picture \ref{rods-images}-a), the particles are spherical in shape. While their diameters is rated to be from 1 to 6~$\mu$m, scanning electron microscopy images of our sample shows that diameters vary roughly from 0.5 to 3~$\mu$m with most of the particles having a diameter around 1 $\mu$m. The particles come in the form of a dry powder and did not undergo specific surface processing. They are injected in the trap using a small metallic tip that is dipped into the powder and brought in the vicinity of the trap. With such micron-sized particles, we can operate the trap with a peak-to-peak voltage ranging from $V_{\rm ac}$=100~V to 4000~V at driving frequencies in the kHz range. Particles are generally injected in the trap under ambient conditions at $V_{\rm ac}$=4000 V and at a trap frequency of a few kHz.

Overall, the imaging set-up (laser, microscope objective, vacuum chamber...) and the trapping conditions are similar to experiments realized with micro-diamonds \cite{vacuumESR,PhysRevLett.121.053602}. However, while levitating diamonds are almost always negatively charged, we found that both positively and negatively charged iron particles can be stably levitating in the trap. This indicates that the tribo-electricity, responsible for the charge acquired by the particles during friction, differs between iron and diamond. Given these observations, on one levitating iron particle, both positive and negative charge patches may be present simultaneously. This is an important observation with regards to the assembly operation that we describe next.

For soft ferromagnets, best angular confinements are obtained for elongated bodies thanks to shape anisotropy \cite{4392562}. In the experiments, we form elongated rods such as the ones shown in Fig.2-c) by levitating simultaneously a few particles in the trap and applying a magnetic field to bind them together using the attractive magnetic forces.

We use the following procedure: 

\begin{itemize}
\item[1-] We inject several iron particles simultaneously in the Paul trap. Given the large size of the trap (200 $\mu$m diameter) compared to the size of the particles, tens of particles can be levitated simultaneously. 

\item[2-] We lower the trap potential and eject some particles using air currents in order to leave 2 to 4 particles in the trap (the optimum aspect ratio for achieving large librational frequencies is calculated next). 

\item[3-] 
We increase the Paul trap confinement back by reducing the trap frequency in order to bring the particles as close as possible from each other. At this stage, particles of the same total charge still repel each other due to electrostatic forces, forming a so-called Coulomb crystal.
 
\item[4-]
We then apply a magnetic field by approaching manually a permanent magnet next to the trap. The magnetic forces between the different particles (which are then magnetized) are attractive. For sufficiently high magnetic fields, on the order of few tens of Gauss, attractive magnetic forces overcome the repulsive electrostatic forces so that the particles bind together, forming a rod aligned in the direction of the magnetic field. The eventual presence of charge patches of different sign on the particles, mentioned in the previous section, might also assist the binding process.
\end{itemize}

Once a rod is formed, particles remain bonded together even when the external magnetic field is nulled, thanks to remanent magnetization and/or due to van-der-Waals forces. 

Fig.\ref{rods-images}-c) shows pictures of several levitating particles obtained by shining incoherent light onto the particles and using the objective to image the particles onto a CCD camera. 
Various shapes can be obtained, which, as we now discuss, will imply different librational confinements.

\subsection{Magnetic confinement of ferromagnetic particles}

We present here theoretical estimations of the torque applied to ferromagnetic particles by an external magnetic field and of the corresponding librational frequency.

\begin{figure}[ht!]
\centerline{\scalebox{0.4}{\includegraphics{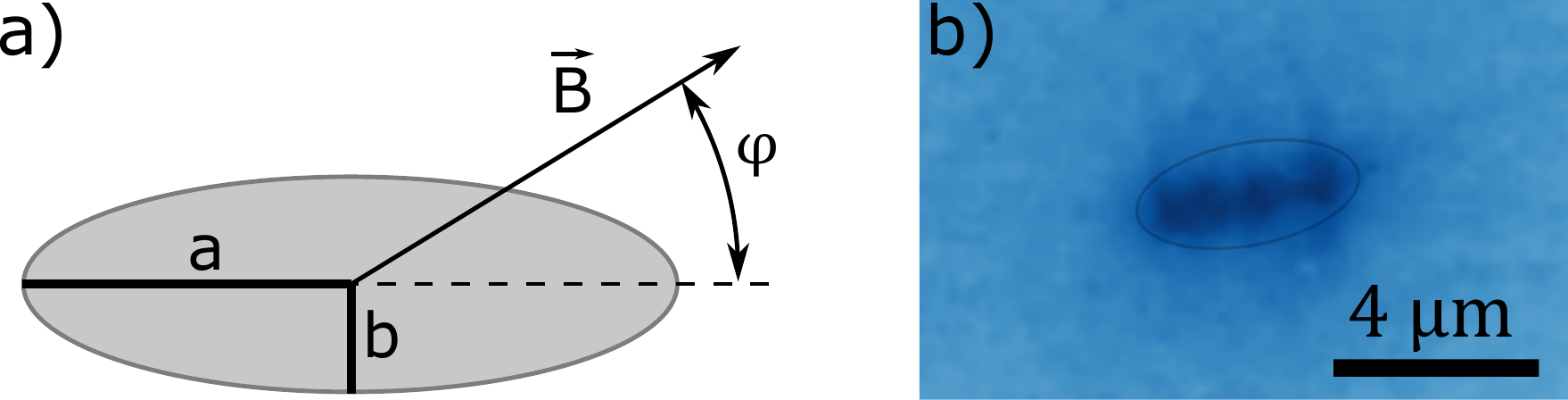}}}
\caption{a) Prolate ellipsoid. b) Image of the particle used for the measurement of the magnetic librational frequency versus magnetic field. An ellipsoid with approximatively the same dimensions is drawn on top of the particle.}
\label{ellipsoids}
\end{figure}

We consider first the torque resulting from shape anisotropy applied on a prolate ellipsoidal soft magnetic body, which tends to align the long axis of the particle along the magnetic field direction (see Fig.\ref{ellipsoids}-a)). In \cite{4392562}, a model to calculate magnetic forces and torques applied on ideal soft ferromagnetic axially symmetrical bodies has been developed and tested experimentally with excellent quantitative agreement. Considering a particle with axial and radial dimensions $2a$ and $2b$, the torque applied on the body by a weak external magnetic field $B$ is given by
\begin{equation}T_{\rm soft}(\phi)=\frac{V (n_r-n_a)}{2\mu_0n_a n_r} B^2 \sin(2\phi),
\end{equation}
where $\phi$ is the angle between the magnetic field and the body symmetry axis, $V=\frac{4\pi}{3}ab^2$ is the volume of the particle, $\mu_0$ is the vacuum magnetic permeability and $n_r$, $n_a \in [0,1]$ are the so-called demagnetisation factors. The latter are purely geometrical factors and can be calculated analytically for ellipsoidal bodies \cite{PhysRev.67.351} to give 
$$n_a = \frac{1}{R^2-1}\left( \frac{R}{2\sqrt{R^2-1}}ln\left( \frac{R+\sqrt{R^2-1}}{R-\sqrt{R^2-1}}\right) -1 \right)$$
and
$$n_r=\frac{1}{2}(1-n_a),$$
where $R=a/b$ is the aspect ratio of the particle. This expression of the torque is valid for a soft ferromagnetic material with large magnetic susceptibility ($\chi \gg 1$) and a magnetic field satisfying
$$ B < \mu_0 m_s\frac{n_a n_r \sqrt{2}}{\sqrt{n_a^2 + n_r^2}},$$
where $m_s$ is the magnetisation at saturation of the material. Under those conditions, the torque does not depend on the magnetic properties of the body but only on its geometry.

From kinematics principles, we find that the librational frequency for the angle $\phi$ resulting from the magnetic torque $T_{\rm soft}$ is given by
\begin{equation}\omega_{\phi} = \sqrt{\frac{V(n_r-n_a)}{I_{\phi}\mu_0n_a n_r} } B,\label{freq}
\end{equation}
where $I_{\phi} = \rho V (a^2 + b^2)/5$ is the relevant component of the particle rotational inertia, $\rho$ being the particle density. 

The librational frequency is proportional to the applied magnetic field and, for a given aspect ratio, it is inversely proportional to the particle size. This latter property can be seen by writing
$$\frac{V}{I_{\phi}}= \frac{1}{V^{\frac{2}{3}}} \frac{5}{\rho}\left( \frac{4\pi}{3} \right)^{\frac{2}{3}} \frac{R^{\frac{2}{3}}}{R^2+1}.$$

\begin{figure}[ht!]
\centerline{\scalebox{0.5}{\includegraphics{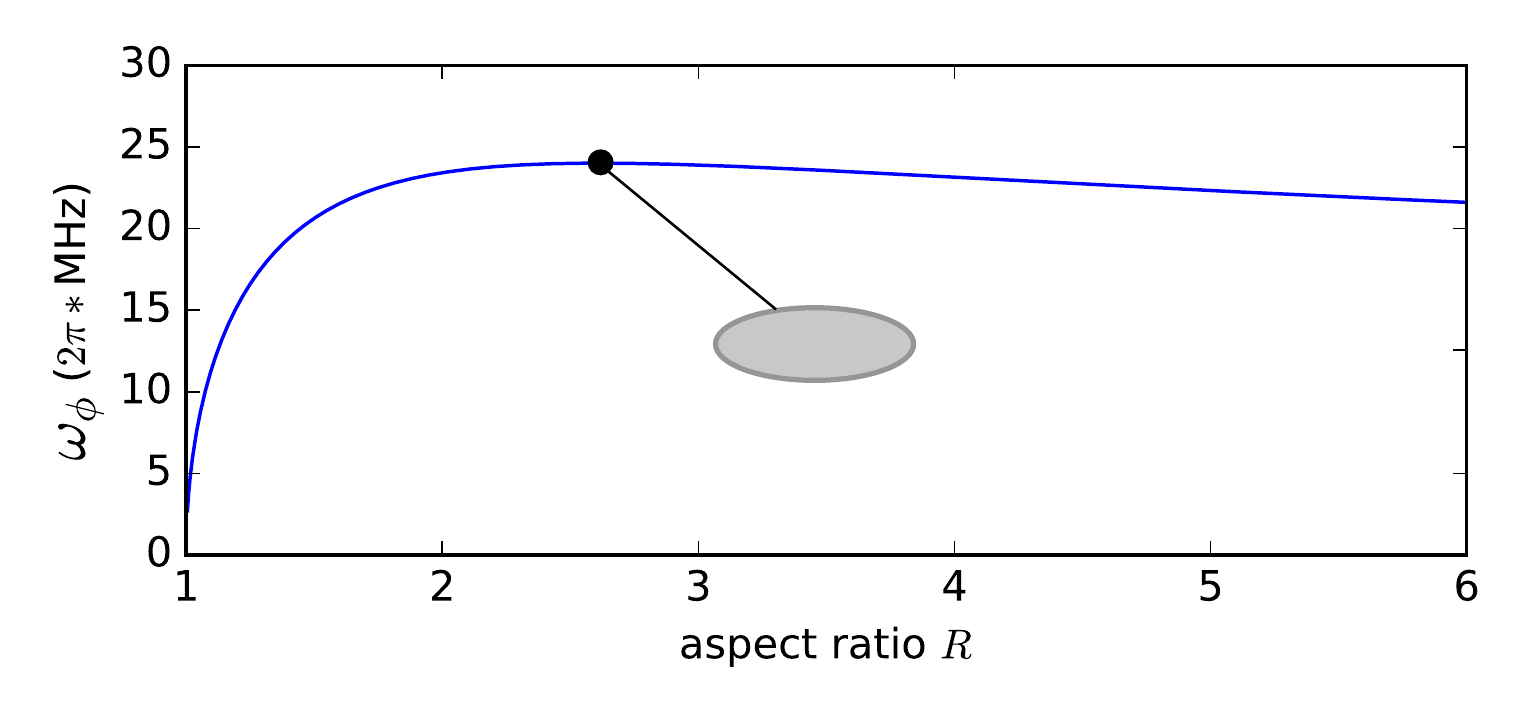}}}
\caption{Magnetic librational frequency versus aspect ratio for a soft ferromagnetic ellipsoid with fixed minor-axis. The ellipse drawn in the middle of the figure has an aspect ratio of $\approx$ 2.6 which gives the highest magnetic libration frequency. The minor-axis is set to 25~nm and the magnetic field to 0.1~T.}
\label{omega-ferro}
\end{figure}

Fig.\ref{omega-ferro} shows $\omega_{\phi}/2\pi$ versus the particle aspect ratio for a fixed minor-axis size. Best librational frequency is obtained for an aspect ratio of $R \approx$ 2.606. For an ellipsoid of 75~nm$\times$25~nm under a field of 0.1~T, we would then obtain $\omega_{\phi} \approx 2\pi\times$24~MHz.

Hard ferromagnetic materials (permanent magnets) could even offer much stronger magnetic librational frequencies than their soft ferromagnetic counterparts. For a magnet with uniform magnetisation $m$ the torque tends to align the particle magnetisation direction along the external magnetic field. As long as the external field does not modify the magnetisation of the material, the torque is simply given by
\begin{equation}
T_{\rm hard}(\phi)= V m B \sin(\phi),
\end{equation}
where $\phi$ is now the angle between the magnetisation and magnetic field directions. This formula holds for any particle shape. Magnetic librational frequencies are then given by $\omega_{\phi} = \sqrt{V m B / I_{\phi}} $. For a 5.4$\mu$m $\times$ 2.5$\mu$m ellipsoidal neodynium particle, under a field of 0.1~T, we find $\omega_{\phi} \approx 2\pi\times$18~MHz. We assume for this a magnetisation of 1.6~T and a density of 7.4 g/cm$^3$. For a smaller 75nm$\times$25nm ellipsoid, we have $\omega_{\phi} \approx 2\pi\times$1.3~GHz.
Experimentally, trapping micro or nano-particles of a hard ferromagnetic material would be highly beneficial to boost mechanical frequencies. 

\section{Detecting the libration of levitating magnets}
 
\begin{figure}
  \centering \scalebox{0.37}{\includegraphics{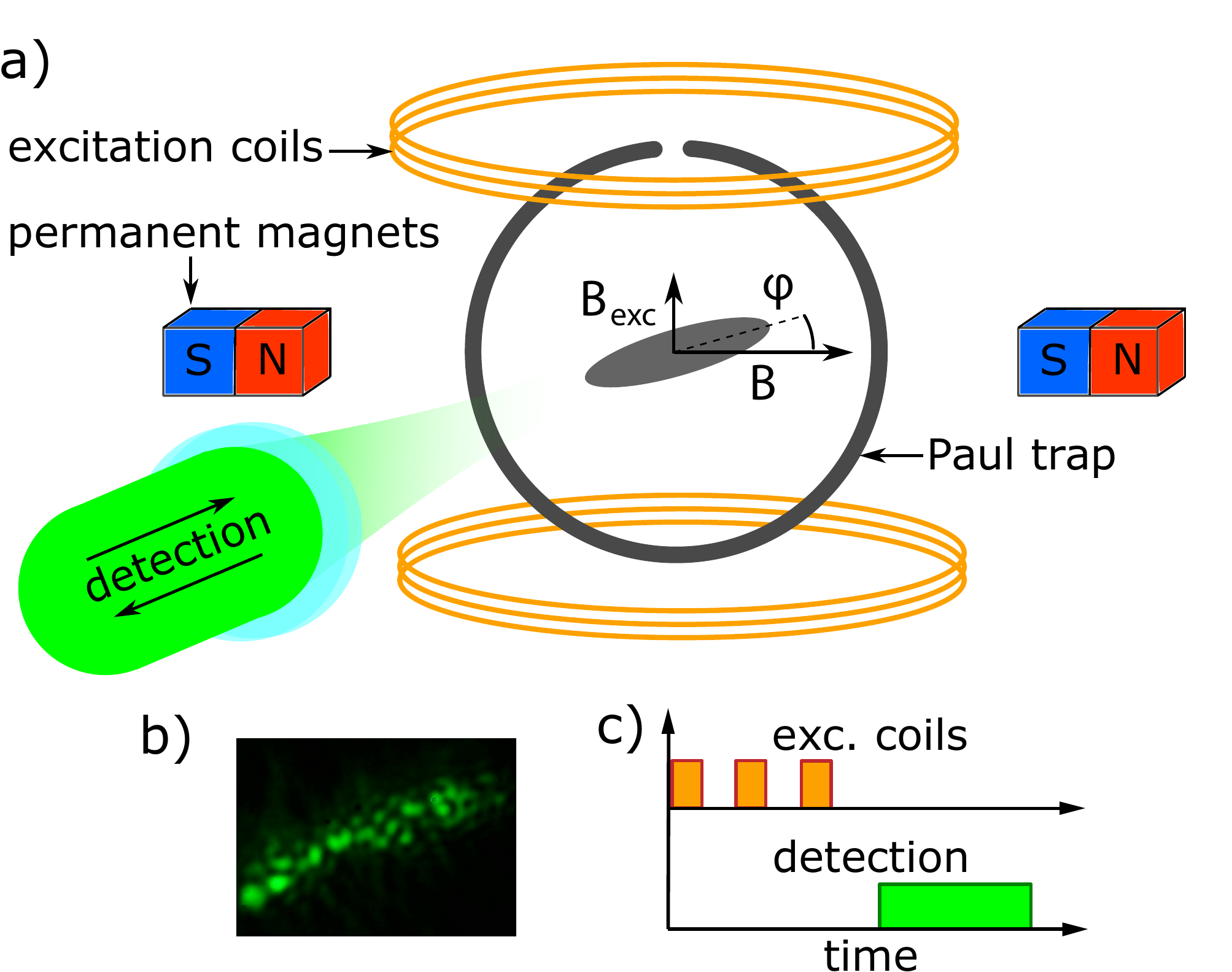}}
  \caption{a) Experimental setup. A ferromagnetic rod is levitating in a ring Paul trap. A pair of permanent magnets generate a uniform magnetic field $\vec B$ that confines the particle orientation. Magnetic coils are used to excite the particle librational motion {\it via} a transverse field $\vec B_{\rm ext}$. b) Image of a levitating ferromagnetic rod showing the speckle pattern upon green laser excitation. c) Sequence used for excitation and ring-down measurements.}\label{setup}
\end{figure}
 
Experimentally, once a rod is formed with a quasi-ellispoidal shape and stably levitating in the Paul trap, we apply a homogeneous magnetic field $B$ of up to 0.1~T using permanent magnets. 
We calibrated the magnetic field generated by the two permanent magnets as a function of their distance from the levitated particle using a Hall probe. 
Then, we use a pair of external coils in Helmholtz configuration in order to excite the librational mode of the levitating ferromagnet. The coils generate a magnetic field perpendicular to the field produced by the permanent magnets used to confine the orientation of the particle, allowing us to displace the particles from their equilibrium orientations. We can thus excite the librational mode by switching the current flowing through the coils. We typically run the coils with a current of 0.5A that is switched off in around 2 $\mu$s using a fast electronic switch (EDR83674/2 from company EDR). While the coils mainly excite the librational mode of motion, the center of mass motion is also weakly excited. To favor the excitation of the librational mode while minimizing the center of mass excitation, the coils current is switched ON and OFF, 3 times, at a frequency close to the librational frequency (See Fig.\ref{setup}-c)). 


After excitation, we measure the particle angle optically. Using the objective inside the vacuum chamber, we focus a green laser onto the particle and collect the reflected light. As shown in Fig.\ref{setup}-b), at the particle image plane which is located few tens of centimeters away, an image of the particle is formed with an additional speckle feature coming from the coherent nature of the illumination. To detect our signal, we focus a small area of this image onto a single-mode optical fibre and detect the photons transmitted through the fibre with a single photon avalanche photodiode. Thanks to the speckle feature, the detected signal is highly sensitive on the particle position and orientation. For a given levitating particle, we can optimize, in real time, the signal coming from the angular displacement of the particle by selecting the most favorable region of the particle image. To do this, we optimize the change in our optical signal while switching the excitation coils at a frequency of 1 Hz. 


Our detection method is not intrinsically linear, i.e. the optical signal is not necessarily linear with the angular displacement, and indeed, harmonics of the librational frequencies can be seen on the signal. However, quasi linearity can be obtained by finely adjusting the detection zone on the particle image.  
\begin{figure}
  \centering \scalebox{0.37}{\includegraphics{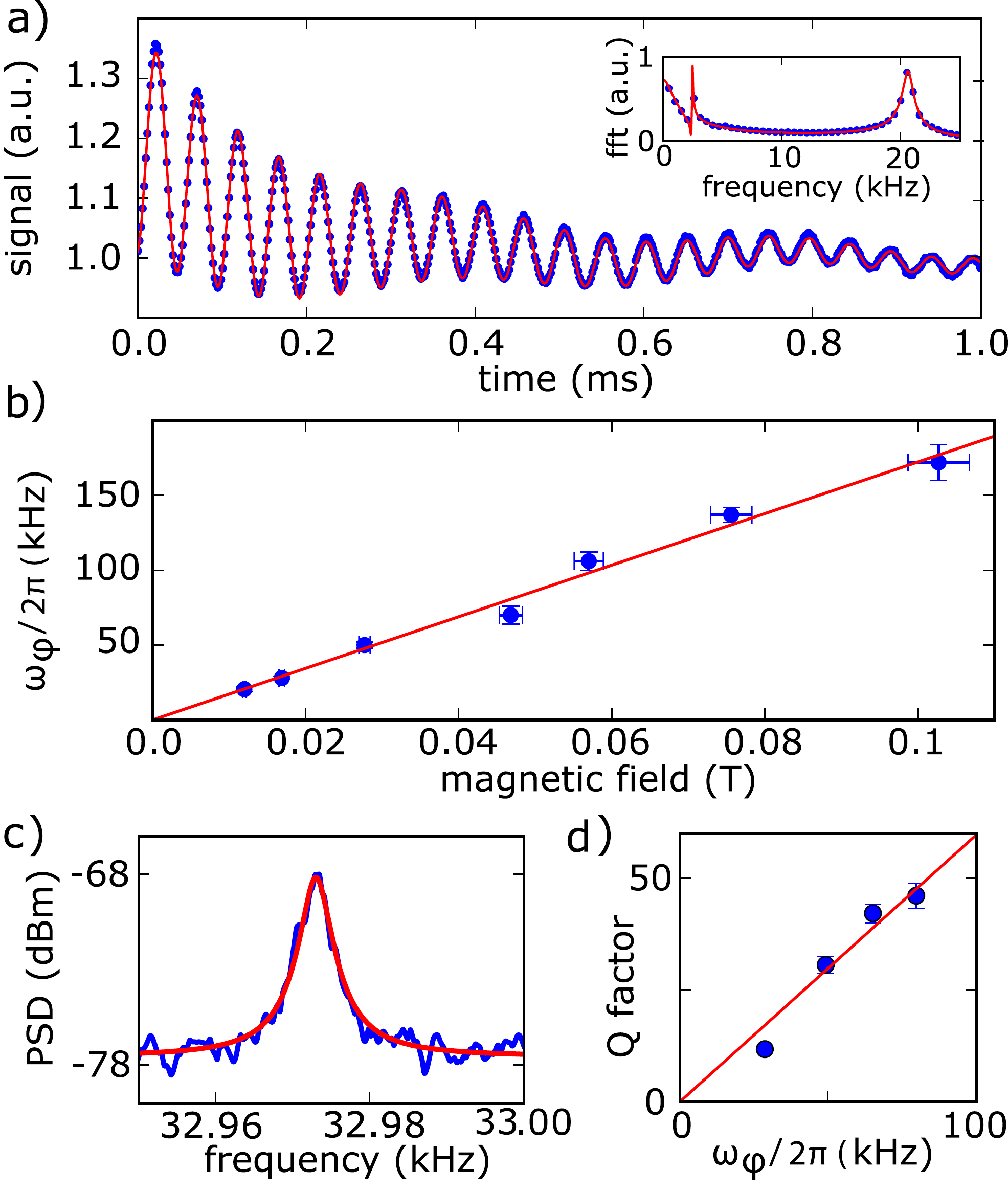}}
  \caption{a) Magnet ring-down under atmospheric pressure. Points are experimental data. The solid line is a fit (explanations in the text). The inset shows the Fourier transform of the temporal signals. 
 b) Librational frequency $\omega_{\phi}/ 2\pi$ versus magnetic field B
c) Power spectrum of the magnet libration undergoing Brownian motion at a pressure of $4.5\times 10^{-2}$~mbar. d) Quality factor at atmospheric pressure as a function of librational frequency. 
.}\label{data}
\end{figure}




\subsection{Libration frequency}

We show in Fig.\ref{data}-a) the optical intensity reflected from the particle under a small magnetic field $B$ of 10~mT as a function of time. As expected, the librational frequency is found at a much higher frequency ($2\pi\times$20.7~kHz) than the maximum Paul trap confinement (given by the trap frequency, here $2\pi\times$2.5~kHz). Residual center of mass excitation is also observed at about 2 kHz, as shown in the inset. 

Fig.\ref{data}-b) shows the librational frequency as a function of the magnetic field. The librational frequency is shown to depend linearly on the applied magnetic field as expected. 
A maximum of $\omega_{\phi}/2\pi=170$~kHz was measured using an external magnetic field close to 0.1~T. 
Approximating our experimentally trapped elongated rods with ellipsoids, we can compare our experimental measurement of $\omega_{\phi}$ to the above theoretical calculations. 
Theoretically, for an ellipsoid of 5.4$\mu$m$\times$2.5$\mu$m as drawn on top of the image, and taking an iron density of 7.86~g/cm$^3$, we find $\omega_{\phi} \approx 2\pi\times$240kHz at a magnetic field of $0.1$~T. This is in fair agreement with the experimental value given the rough approximation made on the particle shape. For this particle shape, the torque equation is valid for magnetic fields below 0.46~T, assuming a magnetization at saturation of iron of 2.2~T. This confirms that a linear dependence of $\omega_{\phi}$ with the magnetic field is effectively expected for the range of fields used in the experiments (from 0 to 0.1~T). Again, since $\omega_{\phi}$ increases with decreasing particle size, larger librational frequencies could be obtained for levitating nano particles. 

\subsection{Quality factor}

In addition to this large trapping frequency, a moderate vacuum of $4.5\times 10^{-2}$ mbar already enables reaching a Q-factor of up to $9.3\pm0.8 \times 10^3$ (see Fig.\ref{data}-c)), and can be even larger under higher vacuum. We could finally verify both the linear dependence of the quality factor as a function of $\omega_{\phi}$ at atmospheric pressure (shown in Fig.\ref{data}-d)).
The current limitation in our experiment is the locked rotation of the magnet around its main axis at lower vacuum pressure, which blurs the speckle pattern we use for detecting the libration. 
This locked rotation, which was also observed with diamond particles \cite{PhysRevLett.121.053602} is a driven rotation of the particle at a frequency equal to half of the Paul trap driving frequency. It will be studied in a forthcoming paper.  \\


The present results show that magnetically confined ferromagnets levitating in a Paul trap form high quality mechanical oscillators, which will find applications in sensitive gyroscopy \cite{Shearwood} and torque balances \cite{PhysRevX.4.021052,Kim2016}.  Compared to other levitation systems, it is also free of the photon shot noise that ultimately limits the quality factor for optical trapped particles \cite{jain2016direct}. Another perspective is the study of the interplay between magnetism and orbital angular momentum \cite{Wang19}, {\it i.e.}. the Einstein-de Haas/Barnett effects \cite{2017PhRvL.119p7202R}, which 
would manifests itself when using particles with much smaller moment of inertia \cite{Kimball}.

Most importantly, the large $\omega_{\phi}/2\pi$ is very promising with regards to coupling the ferromagnet motion to NV centers' spin. It indeed largely exceeds the kHz linewidth of the spin transition in isotopically purified bulk diamond \cite{Maurer1283}, opening the way to the two schemes depicted in Fig. \ref{plat}-a) and b). We now show discuss how to implement the first scheme, and then show experimental results on the way towards the second scheme. 

\section{Coupling levitating magnets to NV centers}

NV centers are point-like defects in diamond that combine both long lifetimes and room temperature read-out and initialization of their electronic spins, making them ideal candidates for an efficient spin-mechanical platform. 
Several schemes have been proposed to couple the electronic spin of an NV center to a magnetized cantilever \cite{Rabl2}, or to the center of mass  \cite{PhysRevA.88.033614,Scala} or librational modes \cite{Ma,delord2017strong} of levitating diamonds. 
For the latter proposals, the NV spin natural quantization axis is essential for applying a spin-dependent torque.
A spin-mechanical coupling can then similarly be achieved using the librational modes of a levitating magnet with the advantage of high motional frequency.

Let us first present the scheme depicted in Fig.\ref{plat}-a), where the librational mode is coupled to a distant NV center in a nano-pyramid at 4K. Here, the NV spin properties in pure bulk diamond at cryogenic temperatures can be fully harnessed: in particular the sharp ESR lines of isotopically enriched diamonds \cite{Maurer1283,Ohno} and the high fidelity spin initialization and read-out \cite{robledo2011high}. 

\subsection{Coupling the librational mode of levitating magnets to a distant NV center : Scheme 1}

We first present the calculations involving the librational mode of levitating magnet coupled to a distant NV center. This corresponds to the proposal a) of the Fig. 1.
Note that this scheme is largely inspired by the proposals to couple an NV spin to a magnetized cantilever \cite{Rabl2} or to the librational mode of a levitating diamond \cite{Ma,delord2017strong}. We therefore only present estimation of the magnitude of the spin-mechanical coupling.

\subsubsection{NV configuration in the field generated by the micro-magnet}

\begin{figure}
\centerline{\scalebox{0.4}{\includegraphics{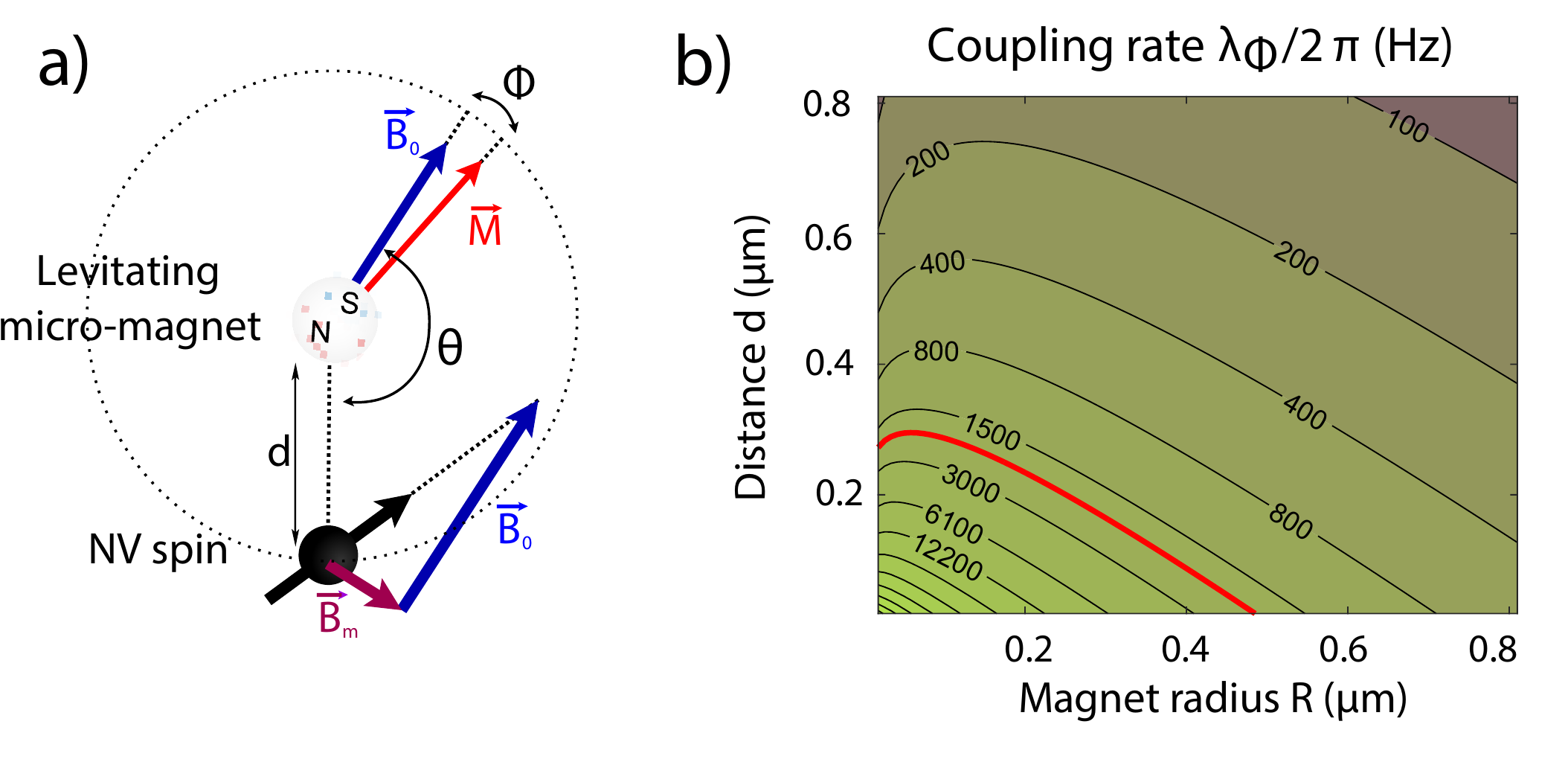}}}
\caption{a) Configuration of the NV center spin relative to the levitating magnet. 
$d$ is the distance between the magnet and the NV center. $\theta$ is the angle between the external magnetic field and the NV-magnet axis and $\phi$ is the angle between the magnetic field and the particle orientation. The NV center orientation is chosen to be along the total field $\mathbf{B_t}=\mathbf{B_0}+\mathbf{B_m}$. b) Coupling rate as a function of particle size and distance from the NV spin using a levitating neodynium micro-magnet with $\omega_\phi/2\pi=200$ kHz. Red curve corresponds to $\lambda_\phi/2\pi=1/500 \mu s$.\label{field_calc}}
\end{figure}

Figure \ref{field_calc}-a) is a schematics of the proposal, including the involved parameters.
The spin-mechanical coupling is obtained through the field $\mathbf{B_m}$ generated by the micro-magnet at the NV position. The amplitude and direction of this magnetic field varies with the angle $\phi$ between the magnetic moment $\mathbf{M}$ of the micro-magnet and its equilibrium position along the external magnetic field $\mathbf{B_0}$. This can then give rise to a coupling between the spin and the magnet libration. 
To obtain a strong enough coupling, the NV center must be placed close to the magnet. 
There are also two other constraints for obtaining a strong coupling:\begin{itemize}
\item The NV center axis must be aligned with the total magnetic field $\mathbf{B_t}=\mathbf{B_0}+\mathbf{B_m}$ to avoid spin state mixing by the transverse magnetic field. This would indeed degrade spin initialization and read-out efficiency \cite{Tetienne2}.
\item The first order variation of $\mathbf{B_m}$ with $\phi$ should be maximal and along the NV axis ${\bf u_{NV}}$.
\end{itemize}

Let us first note that the external magnetic field $\mathbf{B_0}$ orientation and amplitude is already determined as it is needed for confining the angular degree of freedom of the micro-magnet. 
This angle fixes the orientation of the NV axis along the total field $\mathbf{B_t}$. The only degree of freedom left is the angle $\theta$ of the NV around the magnet.

With a spherical levitating magnet, the field generated by the magnet at a point given by the polar coordinates $(d,\theta)$ will be
\begin{equation}
\mathbf{B_m} (\theta)=\frac{M R^3}{3(d+R)^3}\left(2 \cos (\theta-\phi) \mathbf{e_r} + \sin (\theta-\phi) \mathbf{e_\theta} \right),
\end{equation}
where $M$ is the magnetization of the magnet material, $R$ the radius of the magnet.

Since the mean orientation of the magnet is aligned with the external magnetic field $\mathbf{B_0}$ (\textit{i.e.} $\phi=0$) and the NV center is aligned along the total field $\mathbf{B_0}+\mathbf{B_m}$, one can calculate the derivative of the field along the NV axis and perpendicular to it as a function of $\theta$. We find that for a particular angle $\theta_{op}$ the contribution along the NV axis is maximal and the one perpendicular to it cancels. We then have  
\begin{equation}
\theta_{op}=\frac{1}{2} \arccos\left(-\frac{3D_\phi}{6B_0+D_\phi} \right),
\end{equation}
where $D_\phi=M R^3/(d+r)^3$.

At this optimal value we have $\partial \mathbf{B_t} / \partial \phi = D_\phi \mathbf{u_{NV}}$. 
Note that this angle is actually close to $\pi/4$ since the homogeneous magnetic field tends to be stronger than the magnetic field gradient at a large distances from the levitating magnet.

\subsubsection{Spin-mechanical coupling rate}


In the optimum configuration, the total field at the NV center is along the NV axis up to first order in $\phi$ and $D_\phi$. In order to obtain a resonant interaction, a resonant microwave is added to drive the $|m_s=0\rangle \rightarrow |m_s=1\rangle$ spin transition of the NV center electronic spin at a Rabi frequency $\Omega_R=\omega_\phi$. The spin-only part of the Hamiltonian is then diagonal in the $| \pm \rangle=\left( | 0 \rangle\pm | 1\rangle \right) /\sqrt{2}$ basis and the full Hamiltonian can be written :
\begin{equation}
\hat{H}/\hbar=\omega_\phi \hat{S}_z+ \omega_\phi \hat{a}^\dagger \hat{a}+
\gamma D_\phi \phi_0 \hat{S}_x \left( \hat{a}^\dagger+ \hat{a}\right), \label{Hamil2}
\end{equation}
where $\hat{S}_\mu$ are the Pauli matrix operators in the $|\pm \rangle$ basis, $\gamma$ is the gyromagnetic ratio of the NV electron spin, $\hat{a}^\dagger$ and $\hat{a}$ are the creation and annihilation operators for the phonon of the angular motion, and $\phi_0=\sqrt{\hbar/(2 I_\phi \omega_{\phi})}$ its zero-point motion.

The Hamiltonian (\ref{Hamil2}) is a Jaynes-Cummings (JC) Hamiltonian, describing coherent exchange between phonons and the spin at a rate $\lambda_\phi=\gamma D_\phi \phi_0$, which occurs in the so called strong-coupling regime where the decoherence of both the spin and the mechanical oscillator are small compared to $\lambda_\phi/2\pi$.
Fig. \ref{field_calc}-b) shows the coupling strength $\lambda_\phi/2\pi$ as a function of the distance $d$ from the levitating magnet and as a function of the magnet radius $R$. The values of $\lambda_\phi/2\pi$ are calculated assuming a fixed ferromagnet librational frequency $\omega_\phi/2\pi=200$~kHz. The latter assumption ensures that the JC coupling is performed in the resolved sideband regime with ultra-pure CVD-grown diamond tips.
The strong coupling condition $\lambda_\phi/2 \pi>1/T_2^*$ can be reached with $R$ and $d$ in the 100 nm range and considering an NV electronic spin with a coherence lifetime $T_2^* \sim 500~\mu s$ \cite{Maurer1283}. 


\begin{figure}
\centerline{\scalebox{0.5}{\includegraphics{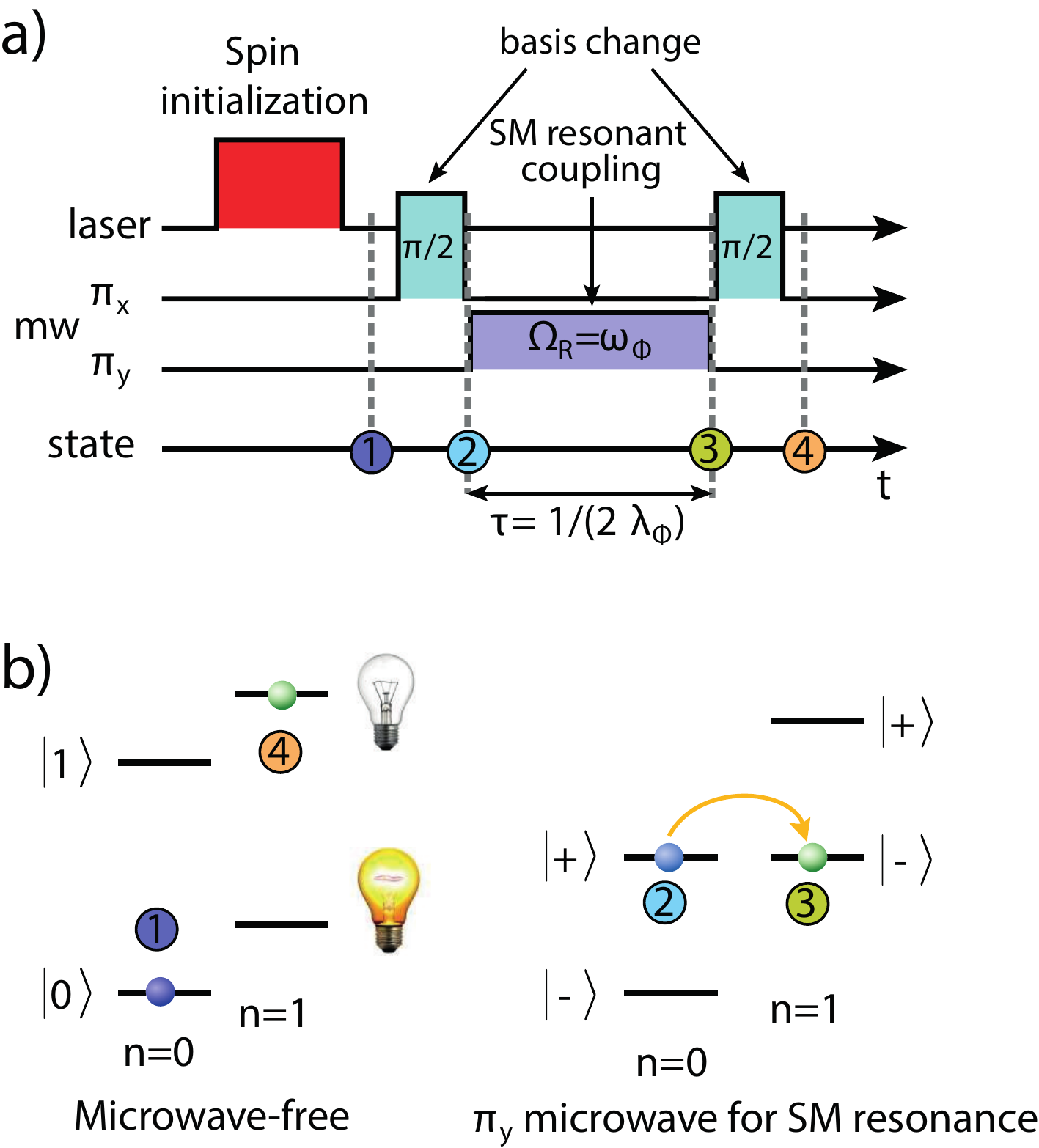}}}
\caption{a) Sequence for preparation of a Fock state $|n=1\rangle$. b) Level diagram for the hybrid system with and without microwave. Colored points show the populated levels during a sequence for creating a $1$ phonon Fock state from the ground state. 
\label{sequence}}
\end{figure}

This coupling scheme and the high degree of coherent control of the spin can be used to generate non classical mechanical states such as the zero phonon ground state or any arbitrary superposition states \cite{PhysRevB.79.041302,law1996arbitraryb}. A key point here is the NV spin excellent initialization fidelity. At cryogenic temperature (4K), it can be initialized in the $|0\rangle$ or $|1\rangle$ state by using resonant optical excitation: after a few optical cycles, the NV spin has a high probability to relax to the spin state for which the optical excitation is non-resonant. Initialization fidelity can then be as high as 0.998 for the $|m_s=\pm 1\rangle$ state \cite{hensen2015loophole}. Efficient ground state spin-cooling can be achieved even in the regime where $\lambda_\phi/2 \pi\ll1/T_2^*$. Generally, in the resolved sideband regime and under high vacuum (in the $10^{-2}$~mbar range), the limitation to efficient cooling is the longitudinal relaxation time $T_1$. Using the proposed scheme will bypass this limit since $T_1(4K)\approx100$~s \cite{Jarmola}.  Using sideband cooling, the mean final phonon number will thus be limited only by the spin initialization fidelity.

Figure \ref{sequence} depicts the sequence that one can use to prepare a one phonon Fock state from the ground state. Here, we assume perfect spin initialization and neglect the decoherences sources. The spin is first initialized in the $|0\rangle$  state and rotated to the $|+\rangle$ state by a $\pi/2$ microwave pulse polarized along the $x$ axis. A resonant microwave is then turned on to achieve the spin-mechanical resonance, with a Rabi frequency $\Omega_R=\omega_\phi$ and a polarization along the $y$ axis (or with a $\pi/2$ dephasing compared to the first microwave pulse). After a time $\tau=1/(2\lambda_\phi)$, the spin is flipped due to coherent exchange with the mechanical oscillator: a subsequent $\pi/2$ pulse ($x$ polarization) maps it unto the $|1\rangle$  state while the phonon number has increased by one. 

In order to first cool the oscillator to the ground state, one only needs to apply this sequence but with an opposite rotation of the spin so that it is in the $|-\rangle$  state before the spin-mechanical coupling. The fidelity of the state obtained in the stationary regime usually depends on the $T_1$ of the oscillator. At low enough pressures, heating from gas molecules will be negligible (see below) so the mean phonon number will here be solely limited by the spin initialization fidelity (0.998) to $\sim$0.002.

Once in the ground state, sequences similar to the one described above can be applied to generate any desired state \cite{Rabl2, LawEberly}. To obtain a high fidelity for these schemes, one is however limited to the strong coupling regime as the decoherence will damp the spin-mechanical Rabi oscillations and result in imperfect pulses.

\subsubsection{Decoherence sources in the scheme 1}

We now discuss the two main decoherence sources that will impact the efficiency of this first cheme.
We first discuss decoherence of the mechanical oscillator and then on the NV center.

{\it Mechanical oscillator} ---The main mechanical decoherence is expected to come from gas collisions with the levitating particle and charges fluctuations. The latter will limit the distance between the NV and the magnet due to fluctuations in the patch potentials on the diamond tip, which will eventually disrupt the charged magnet motion. Estimating the exact influence of such patch potentials is a complicated matter which depends on the surface termination.  
With regards to the former limit, novel theoretical tools are being developed to estimate the librational decoherence rate \cite{stickler2016spatio,zhong2016decoherence}.  The precise value of the decoherence rate will highly depend on the shape of the levitating particle, its roughness and eventually the potential governing the scattering of molecules impacting it. Note that it can be strongly mitigated if one works with an isotropic particle \cite{stickler2016spatio,zhong2016decoherence} or under high enough vacuum.
To give a rough estimate of the heating rate and hence of the lifetime of a mechanical state, we calculate the damping rate of the levitating particle $\Gamma_{\rm gas}$ due to air molecules in the classical regime.

In the Knudsen regime, where the mean free path of the gas molecules is larger than the size of the levitating particle, one gets \cite{fremerey1982spinning} :
\begin{equation}
\Gamma_{\rm gas}=\sigma_{\rm eff} \frac{10 \pi}{R \rho} \frac{P}{\overline{c}},
\end{equation}
with $\sigma_{\rm eff} \sim 1.1$ the accommodation coefficient, $R$ is the radius of the particle, $P$ the pressure in Pascal, $\rho$ its density and $\overline{c}$ the molecular velocity of the considered gas molecules. Using a $1~\mu m$ radius sphere, we find a heating rate of $1$~Hz at $P=10^{-3}$mbar. \\

{\it NV spin decoherence} --- The NV spin itself has an excellent population lifetime ($T_1$) at cryogenic temperatures. It can reach up to a hundred seconds \cite{Jarmola}. However NV spins are typically coupled to the fluctuating nuclear spin bath in the diamond crystal. This results in fluctuations of the frequency of the NV spin and considerably reduces the $T_2^*$. Although this effect can be mitigated using CVD grown diamonds that are isotopically purified to remove $^{13}$C spins and other impurities \cite{mizuochi2009coherence,balasubramanian2009ultralong}, substitutional nitrogen still impacts the linewidth. The smallest observed electron spin linewidth as far as we know was as low as 0.7 kHz ($T_2^*$=460 $\mu s$) \cite{Maurer1283}.
Those coherence times are however obtained with NV center deep deeply buried inside the diamond crystal. When the NV spin is closer to the surface, its properties will suffer from the presence of surface impurities. The magnetic noise from those impurities lowers both the lifetime \cite{Tetienne2} and coherence time \cite{Rosskopf} of shallow NV spins.
It should however be noted that dynamical decoupling could be integrated within a spin-mechanical experiment to solve this issue. 
The sequence described in figure \ref{sequence} is actually similar to a spin-locking decoupling sequence \cite{Ostroff}, with a decoupling rate $\Omega_R=\omega_\phi$. We expect this will protect the NV spin from decoherence caused by the spin bath, as already mentioned in \cite{Rabl2}.
It is difficult to predict the coherence time one can obtain with this scheme, but using a Hahn echo decoupling sequence one can reach coherence time up to 200 $\mu s$ for 5-nm-deep NV spin and of 800 $\mu s$ for 50-nm-deep NV spin \cite{Ohno}.
The impurities themselves could also be driven to decouple them from NV spins \cite{bluvstein}.



\subsection{NV centers in diamonds attached to levitating ferromagnetic particles: Scheme 2}

The other proposed scheme for coupling the ferromagnet librational motion to NV centers is to directly attach diamonds to the ferromagnet. 
This scheme is depicted in Fig. 1-b).
Neglecting the magnetic field produced by the magnet, the interaction Hamiltonian describing the interaction is given by $\hat{H}_{\rm int}=\tilde{\lambda}_\phi \hat{S}_x \left( \hat{a}^\dagger+ \hat{a}\right)$ \cite{delord2017strong}.

Here, $\hat{a}^\dagger$ and $\hat{a}$ are the creation and annihilation operators of the phonon of the librational motion and $\tilde{\lambda}_\phi$ is the spin-mechanical coupling constant
\cite{delord2017strong, Ma}. 
The coupling strength is given by $\lambda_\phi=\gamma B \phi_0$, where $ \phi_0=\sqrt{\frac{\hbar}{2I\omega_\phi}}$, and $I$ is the moment of inertia of the whole composite particle \cite{delord2017strong}.
$\omega_\phi$ is in turn defined by the magnetic field angle and magnitude applied at the location of the hybrid structure, {\it i.e.} by equation (\ref{freq}).
We estimated that with a nanodiamond and a ferromagnet both ellipsoidal with long axes of 80 nm and short axes of 40 nm, a coupling of $\lambda_\phi/2\pi \sim 100$~kHz can be obtained at an external field of 30~mT with an optimum angle of 55$^\circ$ \cite{Ma} while the oscillator reaches a frequency of $\omega_\phi \sim 2\pi\times 3$ MHz.

Obtaining librational frequencies larger than 1 kHz is a relatively straightforward task with levitating hybrid structures containing ferromagnets as opposed to pure-diamonds, where the librational mode frequency depends on dipole or quadrupolar distribution of charges on the diamond surface and are sample dependent. 

Experimentally, we present first steps towards coupling the librational mode of hybrid levitating structures to the electronic spin. 
In one experiment, we prepare particles consisting of 100 nm fluorescent nano-diamonds (FNDs) containing many NV centers attached to the micron-sized iron particles. In another experiment, we evaporate a thin nickel layer on top of diamond particles containing NV centers. Although the confinement frequency of both structures is not as high as with the ferromagnet rods described in the first section, we show that they can levitate stably with librational frequencies above the Paul trap librational confinement.  


\subsubsection{Nanodiamonds attached to levitating magnets}

\begin{figure}[ht!]
\centerline{\scalebox{0.13}{\includegraphics{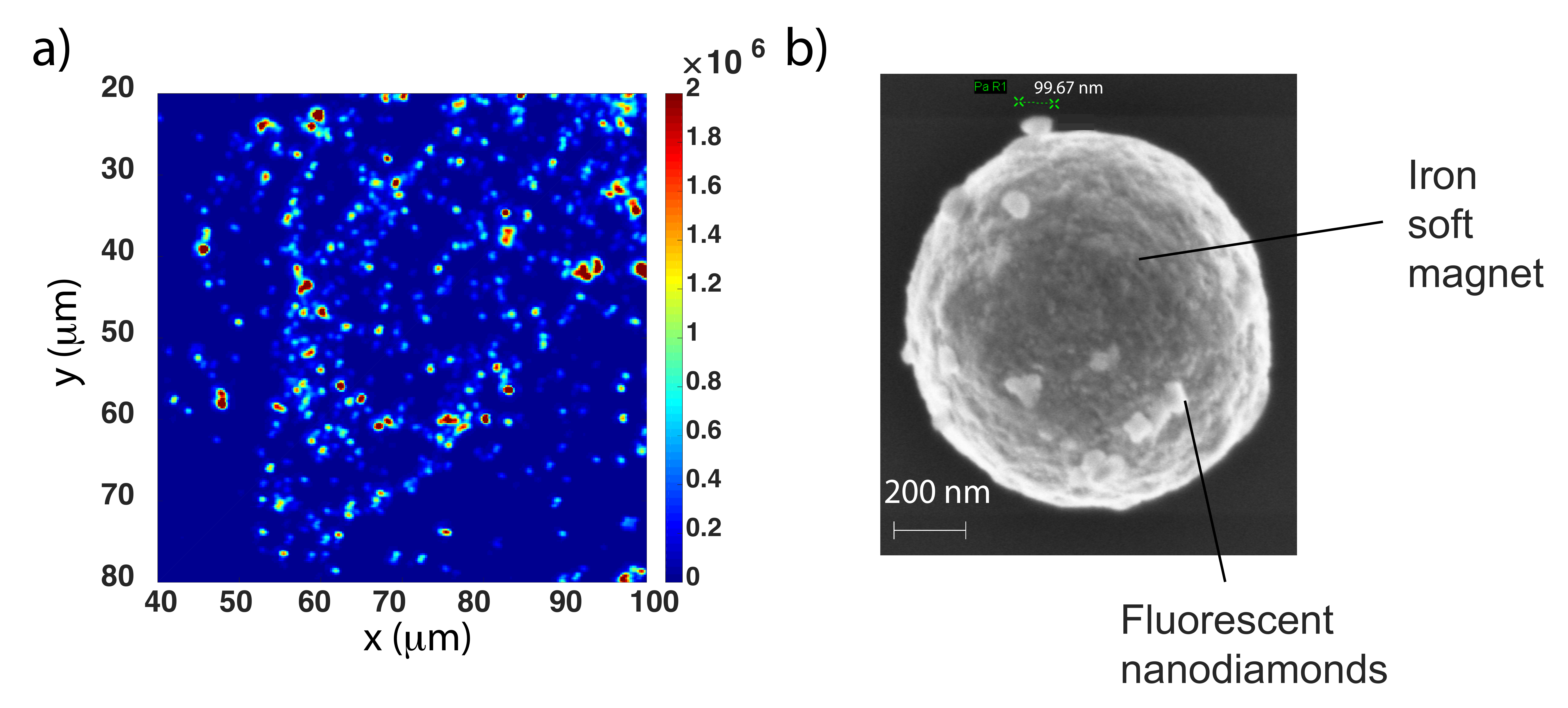}}}
\caption{b) Confocal map of fluorescent nano-diamonds (FND) nebulized on a quartz coverslip. b) Scanning electron microscope image of a hybrid particle composed of Fluorescent Nano-Diamonds (FND) attached to an iron ferromagnet.}
\label{FNDcon}
\end{figure}

We present here the first steps towards the scheme 2 where we use nanodiamonds attached to micromagnets.  
The employed nanodiamonds are bought in the form of a solution containing fluorescent nanodiamonds (brFND-100) from FND biotech, and contain a large fraction of NV centers ($>$1000 NV centers per particle are quoted by the manufacturer).
Fig.\ref{FNDcon}-a) shows a confocal raster scan where the fluorescence from NV centers in FNDs nebulized on a quartz coverslip was recorded as a function of the coverslip position in front of the microscope objective. The latter has a numerical aperture of NA=0.7 here. To attach these NDs to the ferromagnets, we use the following procedure: 

\begin{itemize}
\item[1-] Several iron particles are cast on a clean quartz coverslip. 

\item[2-] A solution containing nanodiamonds (FNDs with 100 nm in diameter) with a large concentration of NV centers in distilled water is prepared and put in the reservoir of ultrasonic nebulizer. The solution is nebulized on top of the quartz coverslip supporting the iron particles. The concentration of FNDs in the nebulizer has been chosen so that single nanodiamonds could be excited optically once in the trap. 

\item[3-] The prepared sample is then scrapped using a small metallic wire and brought close to the trap for loading.
\end{itemize}




\begin{figure}[ht!]
\centerline{\scalebox{0.18}{\includegraphics{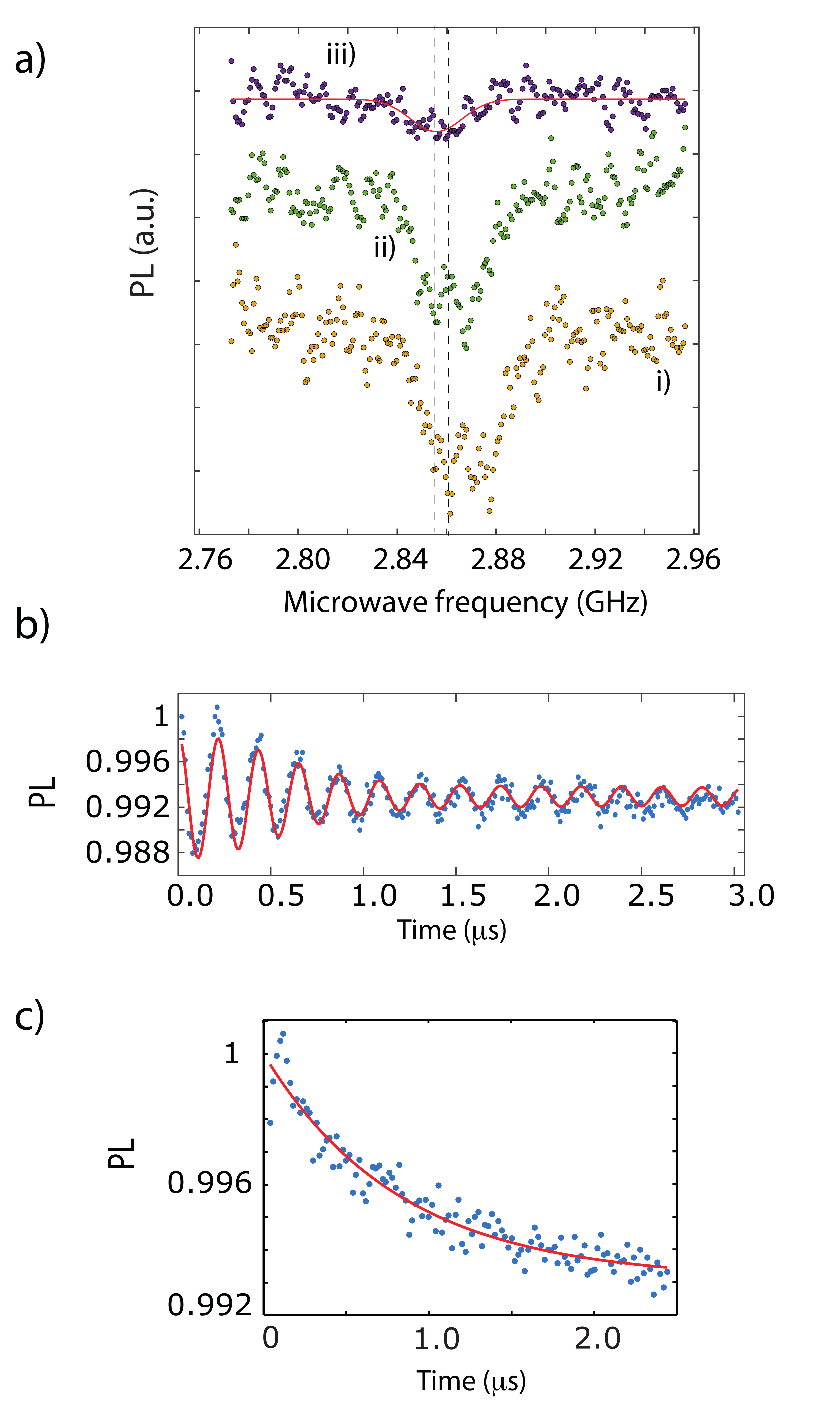}}}
\caption{a) ESR spectrum from NV centers in a nanodiamond attached to a levitating ferromagnet at zero magnetic field. The spectra have been taken at three different vacuum pressures~: P=0.5, 0.07 and 0.01 mbar for traces i) ii) iii) respectively. The experimental data have been offset for clarity.
b) Rabi oscillations for an NV spin transition of the hybrid particle. c) Hahn-echo measurement (T$_{2,\rm echo}$=825 ns).
}\label{Fig1}
\end{figure}

A SEM image of the dried mixture zoomed on one particle is shown in Fig.\ref{FNDcon}-b). 
As shown in Fig.\ref{Fig1}, we could observe Electron-Spin-Resonances (ESR), spin-echoes and Rabi oscillations from the NV centers in the nanodiamonds attached to the ferromagnets, similar to what has been achieved recently with levitating micro-diamonds \cite{PhysRevLett.121.053602}. 

Because of an intersystem crossing in the NV optically excited state, the photoluminescence should drop when the microwave drive the spin at about $D=2.87$~GHz. 
Fig.\ref{Fig1}-a) shows the PL rate from the NV centers as a function of the microwave frequency in this levitating structure. This ESR scan was performed under a pressure of 0.5~mbar for trace i) and in the absence of magnetic field. 
The temperature of the assembly can thus be kept rather low even at significant vacuum levels even when laser light is shone to the particle. 
Trace ii) and trace iii) indicate a drop of the ESR contrast and the shift of the zero field splitting for pressures of P=0.07 and P=0.01 mbars respectively. 
The temperature of the particle can be estimated using a similar procedure than in \cite{vacuumESR} and we found a temperature of around 400~K for trace ii).
Compared to the experiment performed in \cite{vacuumESR}, at least an order of magnitude in pressure could thus be gained, when using similar green laser powers ($100~\mu$W). The reason is likely to be that iron reflects most of the impinging light instead of absorbing it. It could  also be because the ND dissipates heat more efficiently than micro-diamonds due to the larger surface to mass ratio of the NDs compared to the micro-diamonds. 
Fig.\ref{Fig1}-b) shows Rabi oscillations from NV centers in another FND on top of a magnet. It was obtained using one of the NV orientations in the presence of a magnetic field. The measurements were done in the same way as in \cite{PhysRevLett.121.053602}, and here show a decay time of 1.2 $\mu$s at a microwave power of 10 dBm. Fig.\ref{Fig1}-c) shows a Hahn-echo measured on the same spin transition, showing a decay time of 825 ns, slightly shorter than what was observed in \cite{PhysRevLett.121.053602}. 

Let us point out that, although we can show spin-control to a level comparable with our previous work in these hybrid structures, we encounter several issues. 
One point is that, if some ESR lines overlap (which is a problem when one wishes to observe Rabi oscillations and spin-echoes), tuning the external magnetic field orientation with respect to the diamond axis is not enough to lift this degeneracy. 
Indeed, the magnet main axis follows the B field angle, so the NDs that are attached to it are also aligned in the very same way when the B field direction is changed.
Another more important problem that was encountered is that the ESR contrast was often 
reduced compared to the typical values (about 10\% at $B\approx 0~$T). Efficient spin control was thus not possible for all levitating hybrid particles. The presence of charge patches on the iron close to the nano-diamonds could be responsible for this, but further investigations would be needed to nail down this issue. 

\subsubsection{Ferromagnetic coating on a levitating diamond}

\begin{figure}[ht!]
\centerline{\scalebox{0.18}{\includegraphics{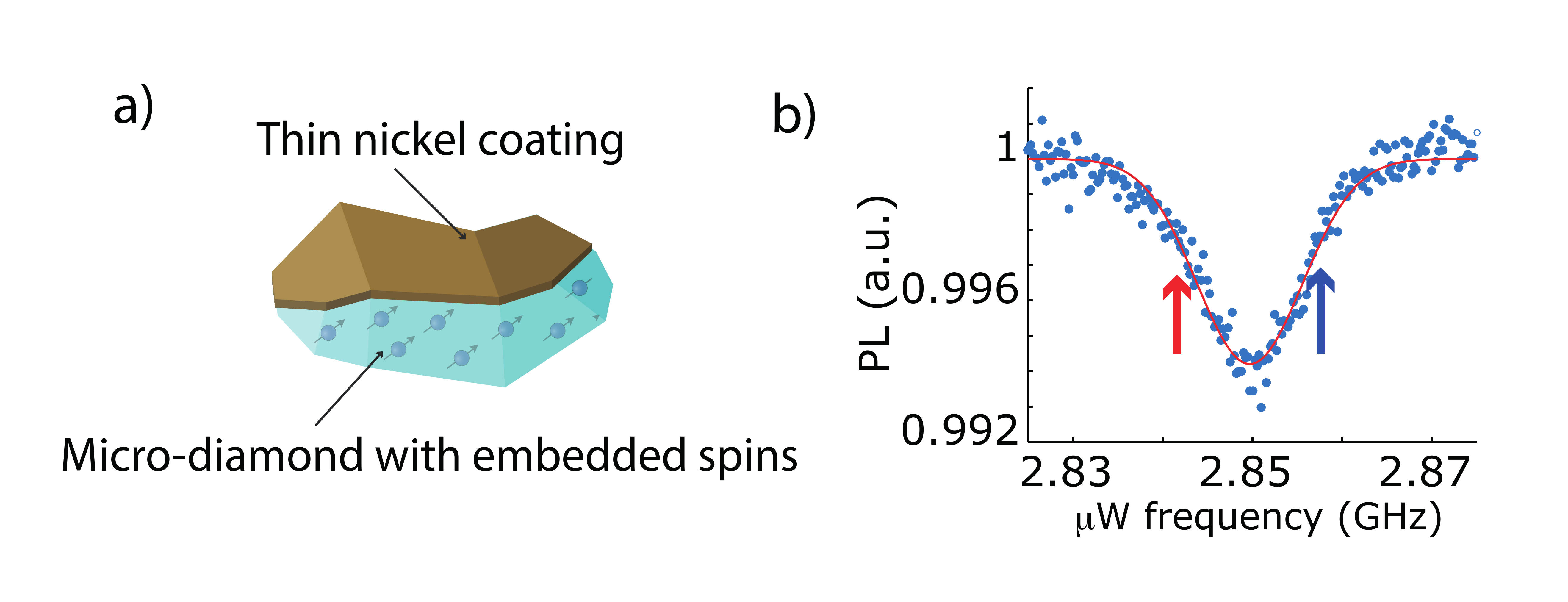}}}
\caption{a) Sketch of a diamond with nickel coating. 
b) ESR spectrum zoomed to one of the four electronic $| m_s=0\rangle  \rightarrow | m_s=-1 \rangle$ transitions, obtained from the levitating hybrid particle.
}\label{Fighybrid2}
\end{figure}

We now turn to the second hybrid structure where a ferromagnet is coated on a diamond. The magnetic coating is also employed here in order to reach larger librational confinement. 
Using this structure, we show spin control and exploit the electronic spin sensitivity to magnetic fields to detect the librational motion.

We produce hybrid particles composed of micro diamonds (MSY 8-12$\mu$m containing around $10^3$ to $10^4$ NV centers, as in \cite{PhysRevLett.121.053602}) with a 200 nm thick magnetic nickel coating on one side of the diamond. This was realized by sputtering nickel atoms from an oven onto a layer of MSY cast on a quartz coverslip. A sketched of the resulting particles is shown in Fig.\ref{Fighybrid2}-a).
The loading of such particles into the trap was done by carefully scratching the coverslip with a 500 microns diameter tip and approaching the tip close to the trap in the absence of external magnetic fields. Doing so, we found similar injection rates than with microdiamonds without the coating and could stably trap the hybrid particle center of mass. Applying a uniform magnetic field of 0.14 T then yields angular confinement of around 4.2~kHz for one of the librational modes, ten times larger than typical librational frequencies provided by the Paul trap \cite{DelordNature}. This mode could be parametrically excited and optically detected using the method described in the first section {\it via} the Helmholtz coils. 

To read out the angular motion of the particle using the NV centers spins embedded in the diamond, we exploit the angular dependency of the NVs ESR transitions frequency. When the orientation of particle changes with respect to the external magnetic field, the frequency of the ESR transition changes and as a result, by applying a microwave field tuned to the side of one ESR transition, the population in the excited spin state then also varies.

The spin state population was read out optically by collecting the NV centers Photo-Luminescence (PL) under green excitation. The diamond side of the particle (without the nickel coating) must be facing the objective so that the laser can excite the NV center and so that the photoluminescence can be extracted efficiently. 
To do so, we maximize the particle photoluminescence while the external magnetic field angle was tuned. Once the diamond side is found, we run the same excitation/detection sequence as for the detection of the magnet motion in the first section, but here monitoring the PL signal and applying a microwave field detuned from one ESR transition.
The measurement is performed for two different microwave detunings, one on each side of the ESR transition where the slope of the ESR signal is maximized. An ESR spectrum from the NV centers is shown in Fig.\ref{Fighybrid2}-b) zoomed to one of the four electronic $| m_s=0\rangle  \rightarrow | m_s=-1 \rangle$ transitions. The two arrows point to the two microwave frequencies that will be used for the libration detection.

\begin{figure}[ht!]
\centerline{\scalebox{0.34}{\includegraphics{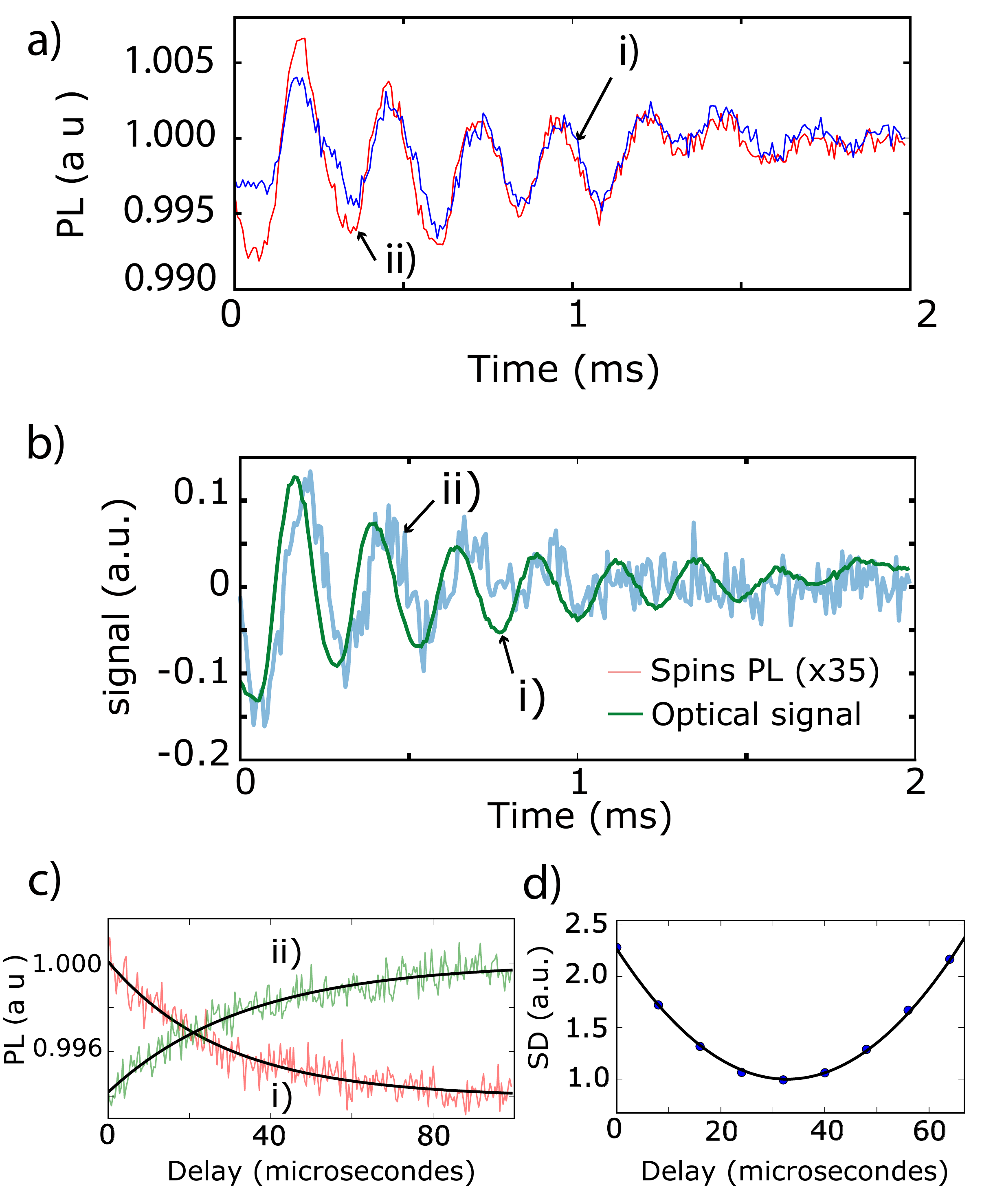}}}
\caption{a) Photo-luminescence signal after parametric excitation of the librational mode for positive (blue) and negative (red) detunings.
b) Librational mode read-out optically (i) and from the spin resonance shift (ii) at frequencies showed by the arrows in Fig. \ref{Fighybrid2}-(b).
c) Temporal evolution of the PL signal after switching ON (trace i) in red) and switching OFF (trace ii), in green) a resonant microwave field. Black solid line are exponential fits to the data. d) Squared difference (SD) between the spin and optical read-out of the motion for various temporal delays between them. Black solid line is a parabolic fit to the data.
}
\label{PLs}
\end{figure}

Fig.\ref{PLs}-a) shows the PL signal as a function of time after parametric excitation of the librational mode. 
Trace i) (in blue online) and trace ii) (in red), correspond to the two signals obtained when the microwave is tuned to the blue and to the red sides of the spin resonance respectively. The amount of PL changes with the particle angular position. This is primarily due to the fact that the green laser illumination conditions changes while the particle oscillates, which then changes the NV laser driving strength and in turn the PL rate. Oscillations that are in phase with the particle motion are visible for both microwave detunings are superimposed with the amplitude modulation coming  solely from the shifts in the spin resonance.
To extract this signal we plot the difference between the two curves.
This allows to remove the spin-independent part of the signal which, again, is the same for both detunings, while keeping the spin-dependent part which has an opposite phase for both detunings.
The resulting PL signal is plotted in Fig.\ref{PLs}-b), trace ii). 
We also measured the variation of the laser signal reflected off the particle as it oscillates, which is shown in trace i), hereafter referred to as direct optical measurements.  
We see that the PL signal reproduces well the direct optical measurement, demonstrating spin read-out of the particle angular motion. We note that, in our experimental conditions, direct optical measurement of the particle motion using the retro-reflected light is much more sensitive than the spin read-out. This is seen in the much better signal-to-noise ratio for direct optical read-out despite a much shorter total acquisition time for the optical (2 mins) than for the spin (50 mins) measurements. 

Interestingly, the spin signal is slightly delayed from the optical one. This can be attributed to the finite response time of the spin population dynamics giving rise to the spin oscillation lagging slightly behind the optical detection, which is in phase with the motion.
To be quantitative, we plot in Fig.\ref{PLs}-c) the PL count rate versus time after switching ON and OFF a resonant microwave field. This allows to measure the rate at which the spins population are excited by the microwave field and polarized by the green light. By fitting the curves with an exponential decay, we found an excitation rate of 28~$\mu$s and a polarization rate of 34~$\mu$s.
Those figures can be compared to the apparent delay between the spin read-out and direct optical measurement of the particle motion.
In Fig.\ref{PLs}-d), we plot the squared difference between the two curves while adding a temporal delay of variable length between them. We found that the two curves coincide best for a delay of 32~$\mu$s which is in perfect agreement with the measured rate of evolution of the spins population.
Note that this phase lag is at the root of cooling in the adiabatic limit \cite{Rabl2,DelordNature}, as in cavity opto-mechanical schemes \cite{Chang19012010}. Optimizing it will thus ensure efficient spin-cooling of this hybrid structure. \\

In conclusion, we proposed methods for quantum mechanical experiments with massive oscillators by coupling levitating magnets to NV centers.
Towards this aim, we levitate magnets in a Paul trap and demonstrate large librational frequencies and Q-factors, which in itself offers the possibility to connect orbital momentum to the spin degree of freedom \cite{2017PhRvL.119p7202R} and to implement ultra-sensitive torque magnetometry \cite{Kim}.
Importantly, it leverages the technical bound on the maximum librational mode frequency that can be attained using a Paul trap angular potential only. We demonstrate that it exceeds the spin transition linewidth of NV centers in CVD grown bulk diamonds, which means that sideband cooling and efficient motional state preparation are within reach.
Last, we demonstrated coherent spin control of NV centers together with read-out of the mechanical motion using the spin of NV centers in hybrid ferromagnet-diamond particles, offering prospects for realizing single-ion-like protocols, such as entangling distant spins using the motional mode of the magnet as a bus \cite{Chen}. \\

\section*{acknowledgements}
We would like to thank Jos\'e Palomo, Christophe Voisin, Nabil Garoum and Andr\'e Thiaville for technical assistance and fruitful discussions. 
GH acknowledges funding by the French National Research Agency (ANR) through the project SMEQUI and by the T-ERC program through the project QUOVADIS.


\begin{thebibliography}{77}%
\makeatletter
\providecommand \@ifxundefined [1]{%
 \@ifx{#1\undefined}
}%
\providecommand \@ifnum [1]{%
 \ifnum #1\expandafter \@firstoftwo
 \else \expandafter \@secondoftwo
 \fi
}%
\providecommand \@ifx [1]{%
 \ifx #1\expandafter \@firstoftwo
 \else \expandafter \@secondoftwo
 \fi
}%
\providecommand \natexlab [1]{#1}%
\providecommand \enquote  [1]{``#1''}%
\providecommand \bibnamefont  [1]{#1}%
\providecommand \bibfnamefont [1]{#1}%
\providecommand \citenamefont [1]{#1}%
\providecommand \href@noop [0]{\@secondoftwo}%
\providecommand \href [0]{\begingroup \@sanitize@url \@href}%
\providecommand \@href[1]{\@@startlink{#1}\@@href}%
\providecommand \@@href[1]{\endgroup#1\@@endlink}%
\providecommand \@sanitize@url [0]{\catcode `\\12\catcode `\$12\catcode
  `\&12\catcode `\#12\catcode `\^12\catcode `\_12\catcode `\%12\relax}%
\providecommand \@@startlink[1]{}%
\providecommand \@@endlink[0]{}%
\providecommand \url  [0]{\begingroup\@sanitize@url \@url }%
\providecommand \@url [1]{\endgroup\@href {#1}{\urlprefix }}%
\providecommand \urlprefix  [0]{URL }%
\providecommand \Eprint [0]{\href }%
\providecommand \doibase [0]{http://dx.doi.org/}%
\providecommand \selectlanguage [0]{\@gobble}%
\providecommand \bibinfo  [0]{\@secondoftwo}%
\providecommand \bibfield  [0]{\@secondoftwo}%
\providecommand \translation [1]{[#1]}%
\providecommand \BibitemOpen [0]{}%
\providecommand \bibitemStop [0]{}%
\providecommand \bibitemNoStop [0]{.\EOS\space}%
\providecommand \EOS [0]{\spacefactor3000\relax}%
\providecommand \BibitemShut  [1]{\csname bibitem#1\endcsname}%
\let\auto@bib@innerbib\@empty
\bibitem [{\citenamefont {Leibfried}\ \emph {et~al.}(2003)\citenamefont
  {Leibfried}, \citenamefont {Blatt}, \citenamefont {Monroe},\ and\
  \citenamefont {Wineland}}]{Leibfried}%
  \BibitemOpen
  \bibfield  {author} {\bibinfo {author} {\bibfnamefont {D.}~\bibnamefont
  {Leibfried}}, \bibinfo {author} {\bibfnamefont {R.}~\bibnamefont {Blatt}},
  \bibinfo {author} {\bibfnamefont {C.}~\bibnamefont {Monroe}}, \ and\ \bibinfo
  {author} {\bibfnamefont {D.}~\bibnamefont {Wineland}},\ }\href@noop {}
  {\bibfield  {journal} {\bibinfo  {journal} {Rev. Mod. Phys.}\ }\textbf
  {\bibinfo {volume} {75}},\ \bibinfo {pages} {281} (\bibinfo {year}
  {2003})}\BibitemShut {NoStop}%
\bibitem [{\citenamefont {Kasevich}\ and\ \citenamefont
  {Chu}(1991)}]{PhysRevLett.67.181}%
  \BibitemOpen
  \bibfield  {author} {\bibinfo {author} {\bibfnamefont {M.}~\bibnamefont
  {Kasevich}}\ and\ \bibinfo {author} {\bibfnamefont {S.}~\bibnamefont {Chu}},\
  }\href {\doibase 10.1103/PhysRevLett.67.181} {\bibfield  {journal} {\bibinfo
  {journal} {Phys. Rev. Lett.}\ }\textbf {\bibinfo {volume} {67}},\ \bibinfo
  {pages} {181} (\bibinfo {year} {1991})}\BibitemShut {NoStop}%
\bibitem [{\citenamefont {Monroe}\ \emph {et~al.}(1996)\citenamefont {Monroe},
  \citenamefont {Meekhof}, \citenamefont {King},\ and\ \citenamefont
  {Wineland}}]{Monroe1131}%
  \BibitemOpen
  \bibfield  {author} {\bibinfo {author} {\bibfnamefont {C.}~\bibnamefont
  {Monroe}}, \bibinfo {author} {\bibfnamefont {D.~M.}\ \bibnamefont {Meekhof}},
  \bibinfo {author} {\bibfnamefont {B.~E.}\ \bibnamefont {King}}, \ and\
  \bibinfo {author} {\bibfnamefont {D.~J.}\ \bibnamefont {Wineland}},\ }\href
  {\doibase 10.1126/science.272.5265.1131} {\ \textbf {\bibinfo {volume}
  {272}},\ \bibinfo {pages} {1131} (\bibinfo {year} {1996})}\BibitemShut
  {NoStop}%
\bibitem [{\citenamefont {Myatt}\ \emph {et~al.}(2000)\citenamefont {Myatt},
  \citenamefont {King}, \citenamefont {Turchette}, \citenamefont {Sackett},
  \citenamefont {Kielpinski}, \citenamefont {Itano}, \citenamefont {Monroe},\
  and\ \citenamefont {Wineland}}]{Myatt2000DecoherenceOQ}%
  \BibitemOpen
  \bibfield  {author} {\bibinfo {author} {\bibfnamefont {C.~J.}\ \bibnamefont
  {Myatt}}, \bibinfo {author} {\bibfnamefont {B.~E.~M.}\ \bibnamefont {King}},
  \bibinfo {author} {\bibfnamefont {Q.~A.}\ \bibnamefont {Turchette}}, \bibinfo
  {author} {\bibfnamefont {C.~A.}\ \bibnamefont {Sackett}}, \bibinfo {author}
  {\bibfnamefont {D.}~\bibnamefont {Kielpinski}}, \bibinfo {author}
  {\bibfnamefont {W.~M.}\ \bibnamefont {Itano}}, \bibinfo {author}
  {\bibfnamefont {C.~A.}\ \bibnamefont {Monroe}}, \ and\ \bibinfo {author}
  {\bibfnamefont {D.~J.}\ \bibnamefont {Wineland}},\ }\href@noop {} {\bibfield
  {journal} {\bibinfo  {journal} {Nature}\ }\textbf {\bibinfo {volume} {403}},\
  \bibinfo {pages} {269} (\bibinfo {year} {2000})}\BibitemShut {NoStop}%
\bibitem [{\citenamefont {Lambert}\ \emph {et~al.}(2008)\citenamefont
  {Lambert}, \citenamefont {Mahboob}, \citenamefont {Pioro-Ladri\`ere},
  \citenamefont {Tokura}, \citenamefont {Tarucha},\ and\ \citenamefont
  {Yamaguchi}}]{PhysRevLett.100.136802}%
  \BibitemOpen
  \bibfield  {author} {\bibinfo {author} {\bibfnamefont {N.}~\bibnamefont
  {Lambert}}, \bibinfo {author} {\bibfnamefont {I.}~\bibnamefont {Mahboob}},
  \bibinfo {author} {\bibfnamefont {M.}~\bibnamefont {Pioro-Ladri\`ere}},
  \bibinfo {author} {\bibfnamefont {Y.}~\bibnamefont {Tokura}}, \bibinfo
  {author} {\bibfnamefont {S.}~\bibnamefont {Tarucha}}, \ and\ \bibinfo
  {author} {\bibfnamefont {H.}~\bibnamefont {Yamaguchi}},\ }\href {\doibase
  10.1103/PhysRevLett.100.136802} {\bibfield  {journal} {\bibinfo  {journal}
  {Phys. Rev. Lett.}\ }\textbf {\bibinfo {volume} {100}},\ \bibinfo {pages}
  {136802} (\bibinfo {year} {2008})}\BibitemShut {NoStop}%
\bibitem [{\citenamefont {Scala}\ \emph {et~al.}(2013)\citenamefont {Scala},
  \citenamefont {Kim}, \citenamefont {Morley}, \citenamefont {Barker},\ and\
  \citenamefont {Bose}}]{Scala}%
  \BibitemOpen
  \bibfield  {author} {\bibinfo {author} {\bibfnamefont {M.}~\bibnamefont
  {Scala}}, \bibinfo {author} {\bibfnamefont {M.~S.}\ \bibnamefont {Kim}},
  \bibinfo {author} {\bibfnamefont {G.~W.}\ \bibnamefont {Morley}}, \bibinfo
  {author} {\bibfnamefont {P.~F.}\ \bibnamefont {Barker}}, \ and\ \bibinfo
  {author} {\bibfnamefont {S.}~\bibnamefont {Bose}},\ }\href@noop {} {\bibfield
   {journal} {\bibinfo  {journal} {Phys. Rev. Lett.}\ }\textbf {\bibinfo
  {volume} {111}},\ \bibinfo {pages} {180403} (\bibinfo {year}
  {2013})}\BibitemShut {NoStop}%
\bibitem [{\citenamefont {Wan}\ \emph {et~al.}(2016)\citenamefont {Wan},
  \citenamefont {Scala}, \citenamefont {Morley}, \citenamefont {Rahman},
  \citenamefont {Ulbricht}, \citenamefont {Bateman}, \citenamefont {Barker},
  \citenamefont {Bose},\ and\ \citenamefont {Kim}}]{Wan}%
  \BibitemOpen
  \bibfield  {author} {\bibinfo {author} {\bibfnamefont {C.}~\bibnamefont
  {Wan}}, \bibinfo {author} {\bibfnamefont {M.}~\bibnamefont {Scala}}, \bibinfo
  {author} {\bibfnamefont {G.~W.}\ \bibnamefont {Morley}}, \bibinfo {author}
  {\bibfnamefont {A.~A.}\ \bibnamefont {Rahman}}, \bibinfo {author}
  {\bibfnamefont {H.}~\bibnamefont {Ulbricht}}, \bibinfo {author}
  {\bibfnamefont {J.}~\bibnamefont {Bateman}}, \bibinfo {author} {\bibfnamefont
  {P.~F.}\ \bibnamefont {Barker}}, \bibinfo {author} {\bibfnamefont
  {S.}~\bibnamefont {Bose}}, \ and\ \bibinfo {author} {\bibfnamefont {M.~S.}\
  \bibnamefont {Kim}},\ }\href@noop {} {\bibfield  {journal} {\bibinfo
  {journal} {Phys. Rev. Lett.}\ }\textbf {\bibinfo {volume} {117}},\ \bibinfo
  {pages} {143003} (\bibinfo {year} {2016})}\BibitemShut {NoStop}%
\bibitem [{\citenamefont {Chen}\ and\ \citenamefont
  {Yin}(2019{\natexlab{a}})}]{chen19}%
  \BibitemOpen
  \bibfield  {author} {\bibinfo {author} {\bibfnamefont {X.-Y.}\ \bibnamefont
  {Chen}}\ and\ \bibinfo {author} {\bibfnamefont {Z.-q.}\ \bibnamefont {Yin}},\
  }\href {\doibase 10.1103/PhysRevA.99.022319} {\bibfield  {journal} {\bibinfo
  {journal} {Phys. Rev. A}\ }\textbf {\bibinfo {volume} {99}},\ \bibinfo
  {pages} {022319} (\bibinfo {year} {2019}{\natexlab{a}})}\BibitemShut
  {NoStop}%
\bibitem [{\citenamefont {Treutlein}\ \emph {et~al.}(2007)\citenamefont
  {Treutlein}, \citenamefont {Hunger}, \citenamefont {Camerer}, \citenamefont
  {H\"ansch},\ and\ \citenamefont {Reichel}}]{PhysRevLett.99.140403}%
  \BibitemOpen
  \bibfield  {author} {\bibinfo {author} {\bibfnamefont {P.}~\bibnamefont
  {Treutlein}}, \bibinfo {author} {\bibfnamefont {D.}~\bibnamefont {Hunger}},
  \bibinfo {author} {\bibfnamefont {S.}~\bibnamefont {Camerer}}, \bibinfo
  {author} {\bibfnamefont {T.~W.}\ \bibnamefont {H\"ansch}}, \ and\ \bibinfo
  {author} {\bibfnamefont {J.}~\bibnamefont {Reichel}},\ }\href {\doibase
  10.1103/PhysRevLett.99.140403} {\bibfield  {journal} {\bibinfo  {journal}
  {Phys. Rev. Lett.}\ }\textbf {\bibinfo {volume} {99}},\ \bibinfo {pages}
  {140403} (\bibinfo {year} {2007})}\BibitemShut {NoStop}%
\bibitem [{\citenamefont {Rabl}\ \emph {et~al.}(2009)\citenamefont {Rabl},
  \citenamefont {Cappellaro}, \citenamefont {Dutt}, \citenamefont {Jiang},
  \citenamefont {Maze},\ and\ \citenamefont {Lukin}}]{PhysRevB.79.041302}%
  \BibitemOpen
  \bibfield  {author} {\bibinfo {author} {\bibfnamefont {P.}~\bibnamefont
  {Rabl}}, \bibinfo {author} {\bibfnamefont {P.}~\bibnamefont {Cappellaro}},
  \bibinfo {author} {\bibfnamefont {M.~V.~G.}\ \bibnamefont {Dutt}}, \bibinfo
  {author} {\bibfnamefont {L.}~\bibnamefont {Jiang}}, \bibinfo {author}
  {\bibfnamefont {J.~R.}\ \bibnamefont {Maze}}, \ and\ \bibinfo {author}
  {\bibfnamefont {M.~D.}\ \bibnamefont {Lukin}},\ }\href@noop {} {\bibfield
  {journal} {\bibinfo  {journal} {Phys. Rev. B}\ }\textbf {\bibinfo {volume}
  {79}},\ \bibinfo {pages} {041302} (\bibinfo {year} {2009})}\BibitemShut
  {NoStop}%
\bibitem [{\citenamefont {{Zhang}}\ \emph {et~al.}(2013)\citenamefont
  {{Zhang}}, \citenamefont {{Zhang}}, \citenamefont {{Zou}}, \citenamefont
  {{Chen}}, \citenamefont {{Yang}}, \citenamefont {{Li}},\ and\ \citenamefont
  {{Feng}}}]{2013OExpr..2129695Z}%
  \BibitemOpen
  \bibfield  {author} {\bibinfo {author} {\bibfnamefont {J.-Q.}\ \bibnamefont
  {{Zhang}}}, \bibinfo {author} {\bibfnamefont {S.}~\bibnamefont {{Zhang}}},
  \bibinfo {author} {\bibfnamefont {J.-H.}\ \bibnamefont {{Zou}}}, \bibinfo
  {author} {\bibfnamefont {L.}~\bibnamefont {{Chen}}}, \bibinfo {author}
  {\bibfnamefont {W.}~\bibnamefont {{Yang}}}, \bibinfo {author} {\bibfnamefont
  {Y.}~\bibnamefont {{Li}}}, \ and\ \bibinfo {author} {\bibfnamefont
  {M.}~\bibnamefont {{Feng}}},\ }\href {\doibase 10.1364/OE.21.029695}
  {\bibfield  {journal} {\bibinfo  {journal} {Optics Express}\ }\textbf
  {\bibinfo {volume} {21}},\ \bibinfo {pages} {29695} (\bibinfo {year}
  {2013})},\ \Eprint {http://arxiv.org/abs/1307.3952} {arXiv:1307.3952
  [quant-ph]} \BibitemShut {NoStop}%
\bibitem [{\citenamefont {Yin}\ \emph {et~al.}(2015)\citenamefont {Yin},
  \citenamefont {Zhao},\ and\ \citenamefont {Li}}]{Yin2015}%
  \BibitemOpen
  \bibfield  {author} {\bibinfo {author} {\bibfnamefont {Z.}~\bibnamefont
  {Yin}}, \bibinfo {author} {\bibfnamefont {N.}~\bibnamefont {Zhao}}, \ and\
  \bibinfo {author} {\bibfnamefont {T.}~\bibnamefont {Li}},\ }\href {\doibase
  10.1007/s11433-015-5651-1} {\bibfield  {journal} {\bibinfo  {journal}
  {Science China Physics, Mechanics {\&} Astronomy}\ }\textbf {\bibinfo
  {volume} {58}},\ \bibinfo {pages} {1} (\bibinfo {year} {2015})}\BibitemShut
  {NoStop}%
\bibitem [{\citenamefont {Li}\ \emph {et~al.}(2016)\citenamefont {Li},
  \citenamefont {Xiang}, \citenamefont {Rabl},\ and\ \citenamefont
  {Nori}}]{PhysRevLett.117.015502}%
  \BibitemOpen
  \bibfield  {author} {\bibinfo {author} {\bibfnamefont {P.-B.}\ \bibnamefont
  {Li}}, \bibinfo {author} {\bibfnamefont {Z.-L.}\ \bibnamefont {Xiang}},
  \bibinfo {author} {\bibfnamefont {P.}~\bibnamefont {Rabl}}, \ and\ \bibinfo
  {author} {\bibfnamefont {F.}~\bibnamefont {Nori}},\ }\href {\doibase
  10.1103/PhysRevLett.117.015502} {\bibfield  {journal} {\bibinfo  {journal}
  {Phys. Rev. Lett.}\ }\textbf {\bibinfo {volume} {117}},\ \bibinfo {pages}
  {015502} (\bibinfo {year} {2016})}\BibitemShut {NoStop}%
\bibitem [{\citenamefont {Delord}\ \emph
  {et~al.}(2017{\natexlab{a}})\citenamefont {Delord}, \citenamefont {Nicolas},
  \citenamefont {Chassagneux},\ and\ \citenamefont
  {H\'etet}}]{delord2017strong}%
  \BibitemOpen
  \bibfield  {author} {\bibinfo {author} {\bibfnamefont {T.}~\bibnamefont
  {Delord}}, \bibinfo {author} {\bibfnamefont {L.}~\bibnamefont {Nicolas}},
  \bibinfo {author} {\bibfnamefont {Y.}~\bibnamefont {Chassagneux}}, \ and\
  \bibinfo {author} {\bibfnamefont {G.}~\bibnamefont {H\'etet}},\ }\href
  {\doibase 10.1103/PhysRevA.96.063810} {\bibfield  {journal} {\bibinfo
  {journal} {Phys. Rev. A}\ }\textbf {\bibinfo {volume} {96}},\ \bibinfo
  {pages} {063810} (\bibinfo {year} {2017}{\natexlab{a}})}\BibitemShut
  {NoStop}%
\bibitem [{\citenamefont {{Macquarrie}}\ \emph {et~al.}(2017)\citenamefont
  {{Macquarrie}}, \citenamefont {{Otten}}, \citenamefont {{Gray}},\ and\
  \citenamefont {{Fuchs}}}]{2017NatCo...814358M}%
  \BibitemOpen
  \bibfield  {author} {\bibinfo {author} {\bibfnamefont {E.~R.}\ \bibnamefont
  {{Macquarrie}}}, \bibinfo {author} {\bibfnamefont {M.}~\bibnamefont
  {{Otten}}}, \bibinfo {author} {\bibfnamefont {S.~K.}\ \bibnamefont {{Gray}}},
  \ and\ \bibinfo {author} {\bibfnamefont {G.~D.}\ \bibnamefont {{Fuchs}}},\
  }\href {\doibase 10.1038/ncomms14358} {\bibfield  {journal} {\bibinfo
  {journal} {Nature Communications}\ }\textbf {\bibinfo {volume} {8}},\
  \bibinfo {eid} {14358} (\bibinfo {year} {2017})},\ \Eprint
  {http://arxiv.org/abs/1605.07131} {arXiv:1605.07131 [quant-ph]} \BibitemShut
  {NoStop}%
\bibitem [{\citenamefont {Abdi}\ \emph {et~al.}(2017)\citenamefont {Abdi},
  \citenamefont {Hwang}, \citenamefont {Aghtar},\ and\ \citenamefont
  {Plenio}}]{Abdi17}%
  \BibitemOpen
  \bibfield  {author} {\bibinfo {author} {\bibfnamefont {M.}~\bibnamefont
  {Abdi}}, \bibinfo {author} {\bibfnamefont {M.-J.}\ \bibnamefont {Hwang}},
  \bibinfo {author} {\bibfnamefont {M.}~\bibnamefont {Aghtar}}, \ and\ \bibinfo
  {author} {\bibfnamefont {M.~B.}\ \bibnamefont {Plenio}},\ }\href {\doibase
  10.1103/PhysRevLett.119.233602} {\bibfield  {journal} {\bibinfo  {journal}
  {Phys. Rev. Lett.}\ }\textbf {\bibinfo {volume} {119}},\ \bibinfo {pages}
  {233602} (\bibinfo {year} {2017})}\BibitemShut {NoStop}%
\bibitem [{\citenamefont {Bassi}\ \emph {et~al.}(2013)\citenamefont {Bassi},
  \citenamefont {Lochan}, \citenamefont {Satin}, \citenamefont {Singh},\ and\
  \citenamefont {Ulbricht}}]{Bassi2013}%
  \BibitemOpen
  \bibfield  {author} {\bibinfo {author} {\bibfnamefont {A.}~\bibnamefont
  {Bassi}}, \bibinfo {author} {\bibfnamefont {K.}~\bibnamefont {Lochan}},
  \bibinfo {author} {\bibfnamefont {S.}~\bibnamefont {Satin}}, \bibinfo
  {author} {\bibfnamefont {T.}~\bibnamefont {Singh}}, \ and\ \bibinfo {author}
  {\bibfnamefont {H.}~\bibnamefont {Ulbricht}},\ }\href {\doibase
  10.1103/RevModPhys.85.471} {\bibfield  {journal} {\bibinfo  {journal} {Rev
  Mod Phys}\ }\textbf {\bibinfo {volume} {85}} (\bibinfo {year} {2013}),\
  10.1103/RevModPhys.85.471}\BibitemShut {NoStop}%
\bibitem [{\citenamefont {Gruber}\ \emph {et~al.}(1997)\citenamefont {Gruber},
  \citenamefont {Drabenstedt}, \citenamefont {Tietz}, \citenamefont {Fleury},
  \citenamefont {Wrachtrup},\ and\ \citenamefont {Borczyskowski}}]{Gruber}%
  \BibitemOpen
  \bibfield  {author} {\bibinfo {author} {\bibfnamefont {A.}~\bibnamefont
  {Gruber}}, \bibinfo {author} {\bibfnamefont {A.}~\bibnamefont {Drabenstedt}},
  \bibinfo {author} {\bibfnamefont {C.}~\bibnamefont {Tietz}}, \bibinfo
  {author} {\bibfnamefont {L.}~\bibnamefont {Fleury}}, \bibinfo {author}
  {\bibfnamefont {J.}~\bibnamefont {Wrachtrup}}, \ and\ \bibinfo {author}
  {\bibfnamefont {C.~v.}\ \bibnamefont {Borczyskowski}},\ }\href {\doibase
  10.1126/science.276.5321.2012} {\bibfield  {journal} {\bibinfo  {journal}
  {Science}\ }\textbf {\bibinfo {volume} {276}},\ \bibinfo {pages} {2012}
  (\bibinfo {year} {1997})}\BibitemShut {NoStop}%
\bibitem [{\citenamefont {Balasubramanian}\ \emph {et~al.}(2009)\citenamefont
  {Balasubramanian}, \citenamefont {Neumann}, \citenamefont {Twitchen},
  \citenamefont {Markham}, \citenamefont {Kolesov}, \citenamefont {Mizuochi},
  \citenamefont {Isoya}, \citenamefont {Achard}, \citenamefont {Beck},
  \citenamefont {Tissler} \emph {et~al.}}]{balasubramanian2009ultralong}%
  \BibitemOpen
  \bibfield  {author} {\bibinfo {author} {\bibfnamefont {G.}~\bibnamefont
  {Balasubramanian}}, \bibinfo {author} {\bibfnamefont {P.}~\bibnamefont
  {Neumann}}, \bibinfo {author} {\bibfnamefont {D.}~\bibnamefont {Twitchen}},
  \bibinfo {author} {\bibfnamefont {M.}~\bibnamefont {Markham}}, \bibinfo
  {author} {\bibfnamefont {R.}~\bibnamefont {Kolesov}}, \bibinfo {author}
  {\bibfnamefont {N.}~\bibnamefont {Mizuochi}}, \bibinfo {author}
  {\bibfnamefont {J.}~\bibnamefont {Isoya}}, \bibinfo {author} {\bibfnamefont
  {J.}~\bibnamefont {Achard}}, \bibinfo {author} {\bibfnamefont
  {J.}~\bibnamefont {Beck}}, \bibinfo {author} {\bibfnamefont {J.}~\bibnamefont
  {Tissler}},  \emph {et~al.},\ }\href@noop {} {\bibfield  {journal} {\bibinfo
  {journal} {Nature materials}\ }\textbf {\bibinfo {volume} {8}},\ \bibinfo
  {pages} {383} (\bibinfo {year} {2009})}\BibitemShut {NoStop}%
\bibitem [{\citenamefont {Maurer}\ \emph {et~al.}(2012)\citenamefont {Maurer},
  \citenamefont {Kucsko}, \citenamefont {Latta}, \citenamefont {Jiang},
  \citenamefont {Yao}, \citenamefont {Bennett}, \citenamefont {Pastawski},
  \citenamefont {Hunger}, \citenamefont {Chisholm}, \citenamefont {Markham},
  \citenamefont {Twitchen}, \citenamefont {Cirac},\ and\ \citenamefont
  {Lukin}}]{Maurer1283}%
  \BibitemOpen
  \bibfield  {author} {\bibinfo {author} {\bibfnamefont {P.~C.}\ \bibnamefont
  {Maurer}}, \bibinfo {author} {\bibfnamefont {G.}~\bibnamefont {Kucsko}},
  \bibinfo {author} {\bibfnamefont {C.}~\bibnamefont {Latta}}, \bibinfo
  {author} {\bibfnamefont {L.}~\bibnamefont {Jiang}}, \bibinfo {author}
  {\bibfnamefont {N.~Y.}\ \bibnamefont {Yao}}, \bibinfo {author} {\bibfnamefont
  {S.~D.}\ \bibnamefont {Bennett}}, \bibinfo {author} {\bibfnamefont
  {F.}~\bibnamefont {Pastawski}}, \bibinfo {author} {\bibfnamefont
  {D.}~\bibnamefont {Hunger}}, \bibinfo {author} {\bibfnamefont
  {N.}~\bibnamefont {Chisholm}}, \bibinfo {author} {\bibfnamefont
  {M.}~\bibnamefont {Markham}}, \bibinfo {author} {\bibfnamefont {D.~J.}\
  \bibnamefont {Twitchen}}, \bibinfo {author} {\bibfnamefont {J.~I.}\
  \bibnamefont {Cirac}}, \ and\ \bibinfo {author} {\bibfnamefont {M.~D.}\
  \bibnamefont {Lukin}},\ }\href {\doibase 10.1126/science.1220513} {\ \textbf
  {\bibinfo {volume} {336}},\ \bibinfo {pages} {1283} (\bibinfo {year}
  {2012})}\BibitemShut {NoStop}%
\bibitem [{\citenamefont {Bar-Gill}\ \emph {et~al.}(2013)\citenamefont
  {Bar-Gill}, \citenamefont {Pham}, \citenamefont {Jarmola}, \citenamefont
  {Budker},\ and\ \citenamefont {Walsworth}}]{BarGill}%
  \BibitemOpen
  \bibfield  {author} {\bibinfo {author} {\bibfnamefont {N.}~\bibnamefont
  {Bar-Gill}}, \bibinfo {author} {\bibfnamefont {L.~M.}\ \bibnamefont {Pham}},
  \bibinfo {author} {\bibfnamefont {A.}~\bibnamefont {Jarmola}}, \bibinfo
  {author} {\bibfnamefont {D.}~\bibnamefont {Budker}}, \ and\ \bibinfo {author}
  {\bibfnamefont {R.~L.}\ \bibnamefont {Walsworth}},\ }\href
  {http://dx.doi.org/10.1038/ncomms2771} {\bibfield  {journal} {\bibinfo
  {journal} {Nature Communications}\ }\textbf {\bibinfo {volume} {4}},\
  \bibinfo {pages} {1743 EP } (\bibinfo {year} {2013})}\BibitemShut {NoStop}%
\bibitem [{\citenamefont {Rabl}(2010)}]{Rabl2}%
  \BibitemOpen
  \bibfield  {author} {\bibinfo {author} {\bibfnamefont {P.}~\bibnamefont
  {Rabl}},\ }\href@noop {} {\bibfield  {journal} {\bibinfo  {journal} {Phys.
  Rev. B}\ }\textbf {\bibinfo {volume} {82}},\ \bibinfo {pages} {165320}
  (\bibinfo {year} {2010})}\BibitemShut {NoStop}%
\bibitem [{\citenamefont {Bennett}\ \emph {et~al.}(2012)\citenamefont
  {Bennett}, \citenamefont {Kolkowitz}, \citenamefont {Unterreithmeier},
  \citenamefont {Rabl}, \citenamefont {Jayich}, \citenamefont {Harris},\ and\
  \citenamefont {Lukin}}]{Bennett}%
  \BibitemOpen
  \bibfield  {author} {\bibinfo {author} {\bibfnamefont {S.~D.}\ \bibnamefont
  {Bennett}}, \bibinfo {author} {\bibfnamefont {S.}~\bibnamefont {Kolkowitz}},
  \bibinfo {author} {\bibfnamefont {Q.~P.}\ \bibnamefont {Unterreithmeier}},
  \bibinfo {author} {\bibfnamefont {P.}~\bibnamefont {Rabl}}, \bibinfo {author}
  {\bibfnamefont {A.~C.~B.}\ \bibnamefont {Jayich}}, \bibinfo {author}
  {\bibfnamefont {J.~G.~E.}\ \bibnamefont {Harris}}, \ and\ \bibinfo {author}
  {\bibfnamefont {M.~D.}\ \bibnamefont {Lukin}},\ }\href@noop {} {\bibfield
  {journal} {\bibinfo  {journal} {New Journal of Physics}\ }\textbf {\bibinfo
  {volume} {14}},\ \bibinfo {pages} {125004} (\bibinfo {year}
  {2012})}\BibitemShut {NoStop}%
\bibitem [{\citenamefont {Kepesidis}\ \emph {et~al.}(2013)\citenamefont
  {Kepesidis}, \citenamefont {Bennett}, \citenamefont {Portolan}, \citenamefont
  {Lukin},\ and\ \citenamefont {Rabl}}]{PhysRevB.88.064105}%
  \BibitemOpen
  \bibfield  {author} {\bibinfo {author} {\bibfnamefont {K.~V.}\ \bibnamefont
  {Kepesidis}}, \bibinfo {author} {\bibfnamefont {S.~D.}\ \bibnamefont
  {Bennett}}, \bibinfo {author} {\bibfnamefont {S.}~\bibnamefont {Portolan}},
  \bibinfo {author} {\bibfnamefont {M.~D.}\ \bibnamefont {Lukin}}, \ and\
  \bibinfo {author} {\bibfnamefont {P.}~\bibnamefont {Rabl}},\ }\href {\doibase
  10.1103/PhysRevB.88.064105} {\bibfield  {journal} {\bibinfo  {journal} {Phys.
  Rev. B}\ }\textbf {\bibinfo {volume} {88}},\ \bibinfo {pages} {064105}
  (\bibinfo {year} {2013})}\BibitemShut {NoStop}%
\bibitem [{\citenamefont {Arcizet}\ \emph {et~al.}(2011)\citenamefont
  {Arcizet}, \citenamefont {Jacques}, \citenamefont {Siria}, \citenamefont
  {Poncharal}, \citenamefont {Vincent},\ and\ \citenamefont
  {Seidelin}}]{Arcizet}%
  \BibitemOpen
  \bibfield  {author} {\bibinfo {author} {\bibfnamefont {O.}~\bibnamefont
  {Arcizet}}, \bibinfo {author} {\bibfnamefont {V.}~\bibnamefont {Jacques}},
  \bibinfo {author} {\bibfnamefont {A.}~\bibnamefont {Siria}}, \bibinfo
  {author} {\bibfnamefont {P.}~\bibnamefont {Poncharal}}, \bibinfo {author}
  {\bibfnamefont {P.}~\bibnamefont {Vincent}}, \ and\ \bibinfo {author}
  {\bibfnamefont {S.}~\bibnamefont {Seidelin}},\ }\href@noop {} {\bibfield
  {journal} {\bibinfo  {journal} {Nat Phys}\ }\textbf {\bibinfo {volume} {7}},\
  \bibinfo {pages} {879} (\bibinfo {year} {2011})}\BibitemShut {NoStop}%
\bibitem [{\citenamefont {Kolkowitz}\ \emph {et~al.}(2012)\citenamefont
  {Kolkowitz}, \citenamefont {Bleszynski~Jayich}, \citenamefont
  {Unterreithmeier}, \citenamefont {Bennett}, \citenamefont {Rabl},
  \citenamefont {Harris},\ and\ \citenamefont {Lukin}}]{Kolkowitz}%
  \BibitemOpen
  \bibfield  {author} {\bibinfo {author} {\bibfnamefont {S.}~\bibnamefont
  {Kolkowitz}}, \bibinfo {author} {\bibfnamefont {A.~C.}\ \bibnamefont
  {Bleszynski~Jayich}}, \bibinfo {author} {\bibfnamefont {Q.~P.}\ \bibnamefont
  {Unterreithmeier}}, \bibinfo {author} {\bibfnamefont {S.~D.}\ \bibnamefont
  {Bennett}}, \bibinfo {author} {\bibfnamefont {P.}~\bibnamefont {Rabl}},
  \bibinfo {author} {\bibfnamefont {J.~G.~E.}\ \bibnamefont {Harris}}, \ and\
  \bibinfo {author} {\bibfnamefont {M.~D.}\ \bibnamefont {Lukin}},\ }\href
  {\doibase 10.1126/science.1216821} {\bibfield  {journal} {\bibinfo  {journal}
  {Science}\ }\textbf {\bibinfo {volume} {335}},\ \bibinfo {pages} {1603}
  (\bibinfo {year} {2012})}\BibitemShut {NoStop}%
\bibitem [{\citenamefont {MacQuarrie}\ \emph {et~al.}(2013)\citenamefont
  {MacQuarrie}, \citenamefont {Gosavi}, \citenamefont {Jungwirth},
  \citenamefont {Bhave},\ and\ \citenamefont {Fuchs}}]{PhysRevLett.111.227602}%
  \BibitemOpen
  \bibfield  {author} {\bibinfo {author} {\bibfnamefont {E.~R.}\ \bibnamefont
  {MacQuarrie}}, \bibinfo {author} {\bibfnamefont {T.~A.}\ \bibnamefont
  {Gosavi}}, \bibinfo {author} {\bibfnamefont {N.~R.}\ \bibnamefont
  {Jungwirth}}, \bibinfo {author} {\bibfnamefont {S.~A.}\ \bibnamefont
  {Bhave}}, \ and\ \bibinfo {author} {\bibfnamefont {G.~D.}\ \bibnamefont
  {Fuchs}},\ }\href {\doibase 10.1103/PhysRevLett.111.227602} {\bibfield
  {journal} {\bibinfo  {journal} {Phys. Rev. Lett.}\ }\textbf {\bibinfo
  {volume} {111}},\ \bibinfo {pages} {227602} (\bibinfo {year}
  {2013})}\BibitemShut {NoStop}%
\bibitem [{\citenamefont {{Teissier}}\ \emph {et~al.}(2014)\citenamefont
  {{Teissier}}, \citenamefont {{Barfuss}}, \citenamefont {{Appel}},
  \citenamefont {{Neu}},\ and\ \citenamefont
  {{Maletinsky}}}]{2014PhRvL.113b0503T}%
  \BibitemOpen
  \bibfield  {author} {\bibinfo {author} {\bibfnamefont {J.}~\bibnamefont
  {{Teissier}}}, \bibinfo {author} {\bibfnamefont {A.}~\bibnamefont
  {{Barfuss}}}, \bibinfo {author} {\bibfnamefont {P.}~\bibnamefont {{Appel}}},
  \bibinfo {author} {\bibfnamefont {E.}~\bibnamefont {{Neu}}}, \ and\ \bibinfo
  {author} {\bibfnamefont {P.}~\bibnamefont {{Maletinsky}}},\ }\href {\doibase
  10.1103/PhysRevLett.113.020503} {\bibfield  {journal} {\bibinfo  {journal}
  {Physical Review Letters}\ }\textbf {\bibinfo {volume} {113}},\ \bibinfo
  {eid} {020503} (\bibinfo {year} {2014})},\ \Eprint
  {http://arxiv.org/abs/1403.3405} {arXiv:1403.3405 [cond-mat.mes-hall]}
  \BibitemShut {NoStop}%
\bibitem [{\citenamefont {Ovartchaiyapong}\ \emph {et~al.}(2014)\citenamefont
  {Ovartchaiyapong}, \citenamefont {Lee}, \citenamefont {Myers},\ and\
  \citenamefont {Jayich}}]{Ovartchaiyapong2014}%
  \BibitemOpen
  \bibfield  {author} {\bibinfo {author} {\bibfnamefont {P.}~\bibnamefont
  {Ovartchaiyapong}}, \bibinfo {author} {\bibfnamefont {K.~W.}\ \bibnamefont
  {Lee}}, \bibinfo {author} {\bibfnamefont {B.~A.}\ \bibnamefont {Myers}}, \
  and\ \bibinfo {author} {\bibfnamefont {A.~C.~B.}\ \bibnamefont {Jayich}},\
  }\href {https://doi.org/10.1038/ncomms5429} {\bibfield  {journal} {\bibinfo
  {journal} {Nature Communications}\ }\textbf {\bibinfo {volume} {5}},\
  \bibinfo {pages} {4429 EP } (\bibinfo {year} {2014})},\ \bibinfo {note}
  {article}\BibitemShut {NoStop}%
\bibitem [{\citenamefont {MacQuarrie}\ \emph {et~al.}(2015)\citenamefont
  {MacQuarrie}, \citenamefont {Gosavi}, \citenamefont {Moehle}, \citenamefont
  {Jungwirth}, \citenamefont {Bhave},\ and\ \citenamefont
  {Fuchs}}]{MacQuarrie15}%
  \BibitemOpen
  \bibfield  {author} {\bibinfo {author} {\bibfnamefont {E.~R.}\ \bibnamefont
  {MacQuarrie}}, \bibinfo {author} {\bibfnamefont {T.~A.}\ \bibnamefont
  {Gosavi}}, \bibinfo {author} {\bibfnamefont {A.~M.}\ \bibnamefont {Moehle}},
  \bibinfo {author} {\bibfnamefont {N.~R.}\ \bibnamefont {Jungwirth}}, \bibinfo
  {author} {\bibfnamefont {S.~A.}\ \bibnamefont {Bhave}}, \ and\ \bibinfo
  {author} {\bibfnamefont {G.~D.}\ \bibnamefont {Fuchs}},\ }\href {\doibase
  10.1364/OPTICA.2.000233} {\bibfield  {journal} {\bibinfo  {journal} {Optica}\
  }\textbf {\bibinfo {volume} {2}},\ \bibinfo {pages} {233} (\bibinfo {year}
  {2015})}\BibitemShut {NoStop}%
\bibitem [{\citenamefont {Barfuss}\ \emph {et~al.}(2015)\citenamefont
  {Barfuss}, \citenamefont {Teissier}, \citenamefont {Neu}, \citenamefont
  {Nunnenkamp},\ and\ \citenamefont {Maletinsky}}]{Barfuss2015}%
  \BibitemOpen
  \bibfield  {author} {\bibinfo {author} {\bibfnamefont {A.}~\bibnamefont
  {Barfuss}}, \bibinfo {author} {\bibfnamefont {J.}~\bibnamefont {Teissier}},
  \bibinfo {author} {\bibfnamefont {E.}~\bibnamefont {Neu}}, \bibinfo {author}
  {\bibfnamefont {A.}~\bibnamefont {Nunnenkamp}}, \ and\ \bibinfo {author}
  {\bibfnamefont {P.}~\bibnamefont {Maletinsky}},\ }\href
  {https://doi.org/10.1038/nphys3411} {\bibfield  {journal} {\bibinfo
  {journal} {Nature Physics}\ }\textbf {\bibinfo {volume} {11}},\ \bibinfo
  {pages} {820 EP } (\bibinfo {year} {2015})}\BibitemShut {NoStop}%
\bibitem [{\citenamefont {{Golter}}\ \emph {et~al.}(2016)\citenamefont
  {{Golter}}, \citenamefont {{Oo}}, \citenamefont {{Amezcua}}, \citenamefont
  {{Stewart}},\ and\ \citenamefont {{Wang}}}]{2016PhRvL.116n3602G}%
  \BibitemOpen
  \bibfield  {author} {\bibinfo {author} {\bibfnamefont {D.~A.}\ \bibnamefont
  {{Golter}}}, \bibinfo {author} {\bibfnamefont {T.}~\bibnamefont {{Oo}}},
  \bibinfo {author} {\bibfnamefont {M.}~\bibnamefont {{Amezcua}}}, \bibinfo
  {author} {\bibfnamefont {K.~A.}\ \bibnamefont {{Stewart}}}, \ and\ \bibinfo
  {author} {\bibfnamefont {H.}~\bibnamefont {{Wang}}},\ }\href {\doibase
  10.1103/PhysRevLett.116.143602} {\bibfield  {journal} {\bibinfo  {journal}
  {Physical Review Letters}\ }\textbf {\bibinfo {volume} {116}},\ \bibinfo
  {eid} {143602} (\bibinfo {year} {2016})},\ \Eprint
  {http://arxiv.org/abs/1603.03804} {arXiv:1603.03804 [quant-ph]} \BibitemShut
  {NoStop}%
\bibitem [{\citenamefont {Yin}\ \emph {et~al.}(2013)\citenamefont {Yin},
  \citenamefont {Li}, \citenamefont {Zhang},\ and\ \citenamefont
  {Duan}}]{PhysRevA.88.033614}%
  \BibitemOpen
  \bibfield  {author} {\bibinfo {author} {\bibfnamefont {Z.-q.}\ \bibnamefont
  {Yin}}, \bibinfo {author} {\bibfnamefont {T.}~\bibnamefont {Li}}, \bibinfo
  {author} {\bibfnamefont {X.}~\bibnamefont {Zhang}}, \ and\ \bibinfo {author}
  {\bibfnamefont {L.~M.}\ \bibnamefont {Duan}},\ }\href {\doibase
  10.1103/PhysRevA.88.033614} {\bibfield  {journal} {\bibinfo  {journal} {Phys.
  Rev. A}\ }\textbf {\bibinfo {volume} {88}},\ \bibinfo {pages} {033614}
  (\bibinfo {year} {2013})}\BibitemShut {NoStop}%
\bibitem [{\citenamefont {Ma}\ \emph {et~al.}(2017)\citenamefont {Ma},
  \citenamefont {Hoang}, \citenamefont {Gong}, \citenamefont {Li},\ and\
  \citenamefont {Yin}}]{Ma}%
  \BibitemOpen
  \bibfield  {author} {\bibinfo {author} {\bibfnamefont {Y.}~\bibnamefont
  {Ma}}, \bibinfo {author} {\bibfnamefont {T.~M.}\ \bibnamefont {Hoang}},
  \bibinfo {author} {\bibfnamefont {M.}~\bibnamefont {Gong}}, \bibinfo {author}
  {\bibfnamefont {T.}~\bibnamefont {Li}}, \ and\ \bibinfo {author}
  {\bibfnamefont {Z.-q.}\ \bibnamefont {Yin}},\ }\href {\doibase
  10.1103/PhysRevA.96.023827} {\bibfield  {journal} {\bibinfo  {journal} {Phys.
  Rev. A}\ }\textbf {\bibinfo {volume} {96}},\ \bibinfo {pages} {023827}
  (\bibinfo {year} {2017})}\BibitemShut {NoStop}%
\bibitem [{\citenamefont {Chang}\ \emph {et~al.}(2010)\citenamefont {Chang},
  \citenamefont {Regal}, \citenamefont {Papp}, \citenamefont {Wilson},
  \citenamefont {Ye}, \citenamefont {Painter}, \citenamefont {Kimble},\ and\
  \citenamefont {Zoller}}]{Chang19012010}%
  \BibitemOpen
  \bibfield  {author} {\bibinfo {author} {\bibfnamefont {D.~E.}\ \bibnamefont
  {Chang}}, \bibinfo {author} {\bibfnamefont {C.~A.}\ \bibnamefont {Regal}},
  \bibinfo {author} {\bibfnamefont {S.~B.}\ \bibnamefont {Papp}}, \bibinfo
  {author} {\bibfnamefont {D.~J.}\ \bibnamefont {Wilson}}, \bibinfo {author}
  {\bibfnamefont {J.}~\bibnamefont {Ye}}, \bibinfo {author} {\bibfnamefont
  {O.}~\bibnamefont {Painter}}, \bibinfo {author} {\bibfnamefont {H.~J.}\
  \bibnamefont {Kimble}}, \ and\ \bibinfo {author} {\bibfnamefont
  {P.}~\bibnamefont {Zoller}},\ }\href {\doibase 10.1073/pnas.0912969107}
  {\bibfield  {journal} {\bibinfo  {journal} {Proceedings of the National
  Academy of Sciences}\ }\textbf {\bibinfo {volume} {107}},\ \bibinfo {pages}
  {1005} (\bibinfo {year} {2010})}\BibitemShut {NoStop}%
\bibitem [{\citenamefont {Romero-Isart}\ \emph {et~al.}(2011)\citenamefont
  {Romero-Isart}, \citenamefont {Pflanzer}, \citenamefont {Juan}, \citenamefont
  {Quidant}, \citenamefont {Kiesel}, \citenamefont {Aspelmeyer},\ and\
  \citenamefont {Cirac}}]{Romero2}%
  \BibitemOpen
  \bibfield  {author} {\bibinfo {author} {\bibfnamefont {O.}~\bibnamefont
  {Romero-Isart}}, \bibinfo {author} {\bibfnamefont {A.~C.}\ \bibnamefont
  {Pflanzer}}, \bibinfo {author} {\bibfnamefont {M.~L.}\ \bibnamefont {Juan}},
  \bibinfo {author} {\bibfnamefont {R.}~\bibnamefont {Quidant}}, \bibinfo
  {author} {\bibfnamefont {N.}~\bibnamefont {Kiesel}}, \bibinfo {author}
  {\bibfnamefont {M.}~\bibnamefont {Aspelmeyer}}, \ and\ \bibinfo {author}
  {\bibfnamefont {J.~I.}\ \bibnamefont {Cirac}},\ }\href@noop {} {\bibfield
  {journal} {\bibinfo  {journal} {Phys. Rev. A}\ }\textbf {\bibinfo {volume}
  {83}},\ \bibinfo {pages} {013803} (\bibinfo {year} {2011})}\BibitemShut
  {NoStop}%
\bibitem [{\citenamefont {Neukirch}\ \emph {et~al.}(2013)\citenamefont
  {Neukirch}, \citenamefont {Gieseler}, \citenamefont {Quidant}, \citenamefont
  {Novotny},\ and\ \citenamefont {Nick~Vamivakas}}]{Neukirch}%
  \BibitemOpen
  \bibfield  {author} {\bibinfo {author} {\bibfnamefont {L.~P.}\ \bibnamefont
  {Neukirch}}, \bibinfo {author} {\bibfnamefont {J.}~\bibnamefont {Gieseler}},
  \bibinfo {author} {\bibfnamefont {R.}~\bibnamefont {Quidant}}, \bibinfo
  {author} {\bibfnamefont {L.}~\bibnamefont {Novotny}}, \ and\ \bibinfo
  {author} {\bibfnamefont {A.}~\bibnamefont {Nick~Vamivakas}},\ }\href@noop {}
  {\bibfield  {journal} {\bibinfo  {journal} {Optics Letters}\ }\textbf
  {\bibinfo {volume} {38}},\ \bibinfo {pages} {2976} (\bibinfo {year}
  {2013})}\BibitemShut {NoStop}%
\bibitem [{\citenamefont {Hoang}\ \emph
  {et~al.}(2016{\natexlab{a}})\citenamefont {Hoang}, \citenamefont {Ahn},
  \citenamefont {Bang},\ and\ \citenamefont {Li}}]{Hoang}%
  \BibitemOpen
  \bibfield  {author} {\bibinfo {author} {\bibfnamefont {T.~M.}\ \bibnamefont
  {Hoang}}, \bibinfo {author} {\bibfnamefont {J.}~\bibnamefont {Ahn}}, \bibinfo
  {author} {\bibfnamefont {J.}~\bibnamefont {Bang}}, \ and\ \bibinfo {author}
  {\bibfnamefont {T.}~\bibnamefont {Li}},\ }\href@noop {} {\bibfield  {journal}
  {\bibinfo  {journal} {Nature Communications}\ }\textbf {\bibinfo {volume}
  {7}},\ \bibinfo {pages} {12250 EP } (\bibinfo {year}
  {2016}{\natexlab{a}})}\BibitemShut {NoStop}%
\bibitem [{\citenamefont {Hoang}\ \emph
  {et~al.}(2016{\natexlab{b}})\citenamefont {Hoang}, \citenamefont {Ma},
  \citenamefont {Ahn}, \citenamefont {Bang}, \citenamefont {Robicheaux},
  \citenamefont {Yin},\ and\ \citenamefont {Li}}]{PhysRevLett.117.123604}%
  \BibitemOpen
  \bibfield  {author} {\bibinfo {author} {\bibfnamefont {T.~M.}\ \bibnamefont
  {Hoang}}, \bibinfo {author} {\bibfnamefont {Y.}~\bibnamefont {Ma}}, \bibinfo
  {author} {\bibfnamefont {J.}~\bibnamefont {Ahn}}, \bibinfo {author}
  {\bibfnamefont {J.}~\bibnamefont {Bang}}, \bibinfo {author} {\bibfnamefont
  {F.}~\bibnamefont {Robicheaux}}, \bibinfo {author} {\bibfnamefont {Z.-Q.}\
  \bibnamefont {Yin}}, \ and\ \bibinfo {author} {\bibfnamefont
  {T.}~\bibnamefont {Li}},\ }\href {\doibase 10.1103/PhysRevLett.117.123604}
  {\bibfield  {journal} {\bibinfo  {journal} {Phys. Rev. Lett.}\ }\textbf
  {\bibinfo {volume} {117}},\ \bibinfo {pages} {123604} (\bibinfo {year}
  {2016}{\natexlab{b}})}\BibitemShut {NoStop}%
\bibitem [{\citenamefont {Rahman}\ \emph {et~al.}(2016)\citenamefont {Rahman},
  \citenamefont {Frangeskou}, \citenamefont {Kim}, \citenamefont {Bose},
  \citenamefont {Morley},\ and\ \citenamefont {Barker}}]{Rahman}%
  \BibitemOpen
  \bibfield  {author} {\bibinfo {author} {\bibfnamefont {A.~T. M.~A.}\
  \bibnamefont {Rahman}}, \bibinfo {author} {\bibfnamefont {A.~C.}\
  \bibnamefont {Frangeskou}}, \bibinfo {author} {\bibfnamefont {M.~S.}\
  \bibnamefont {Kim}}, \bibinfo {author} {\bibfnamefont {S.}~\bibnamefont
  {Bose}}, \bibinfo {author} {\bibfnamefont {G.~W.}\ \bibnamefont {Morley}}, \
  and\ \bibinfo {author} {\bibfnamefont {P.~F.}\ \bibnamefont {Barker}},\
  }\href@noop {} {\bibfield  {journal} {\bibinfo  {journal} {Scientific
  Reports}\ }\textbf {\bibinfo {volume} {6}},\ \bibinfo {pages} {21633 EP }
  (\bibinfo {year} {2016})}\BibitemShut {NoStop}%
\bibitem [{\citenamefont {Hsu}\ \emph {et~al.}(2016)\citenamefont {Hsu},
  \citenamefont {Ji}, \citenamefont {Lewandowski},\ and\ \citenamefont
  {D'Urso}}]{Hsu}%
  \BibitemOpen
  \bibfield  {author} {\bibinfo {author} {\bibfnamefont {J.-F.}\ \bibnamefont
  {Hsu}}, \bibinfo {author} {\bibfnamefont {P.}~\bibnamefont {Ji}}, \bibinfo
  {author} {\bibfnamefont {C.~W.}\ \bibnamefont {Lewandowski}}, \ and\ \bibinfo
  {author} {\bibfnamefont {B.}~\bibnamefont {D'Urso}},\ }\href@noop {}
  {\bibfield  {journal} {\bibinfo  {journal} {Scientific Reports}\ }\textbf
  {\bibinfo {volume} {6}},\ \bibinfo {pages} {30125 EP } (\bibinfo {year}
  {2016})}\BibitemShut {NoStop}%
\bibitem [{\citenamefont {Kuhlicke}\ \emph {et~al.}(2014)\citenamefont
  {Kuhlicke}, \citenamefont {Schell}, \citenamefont {Zoll},\ and\ \citenamefont
  {Benson}}]{Kuhlicke}%
  \BibitemOpen
  \bibfield  {author} {\bibinfo {author} {\bibfnamefont {A.}~\bibnamefont
  {Kuhlicke}}, \bibinfo {author} {\bibfnamefont {A.~W.}\ \bibnamefont
  {Schell}}, \bibinfo {author} {\bibfnamefont {J.}~\bibnamefont {Zoll}}, \ and\
  \bibinfo {author} {\bibfnamefont {O.}~\bibnamefont {Benson}},\ }\href@noop {}
  {\bibfield  {journal} {\bibinfo  {journal} {Applied Physics Letters}\
  }\textbf {\bibinfo {volume} {105}} (\bibinfo {year} {2014})}\BibitemShut
  {NoStop}%
\bibitem [{\citenamefont {Delord}\ \emph
  {et~al.}(2017{\natexlab{b}})\citenamefont {Delord}, \citenamefont {Nicolas},
  \citenamefont {Schwab},\ and\ \citenamefont {H{\'e}tet}}]{delord2016}%
  \BibitemOpen
  \bibfield  {author} {\bibinfo {author} {\bibfnamefont {T.}~\bibnamefont
  {Delord}}, \bibinfo {author} {\bibfnamefont {L.}~\bibnamefont {Nicolas}},
  \bibinfo {author} {\bibfnamefont {L.}~\bibnamefont {Schwab}}, \ and\ \bibinfo
  {author} {\bibfnamefont {G.}~\bibnamefont {H{\'e}tet}},\ }\href@noop {}
  {\bibfield  {journal} {\bibinfo  {journal} {New Journal of Physics}\ }\textbf
  {\bibinfo {volume} {19}},\ \bibinfo {pages} {033031} (\bibinfo {year}
  {2017}{\natexlab{b}})}\BibitemShut {NoStop}%
\bibitem [{\citenamefont {Delord}\ \emph {et~al.}(2018)\citenamefont {Delord},
  \citenamefont {Huillery}, \citenamefont {Schwab}, \citenamefont {Nicolas},
  \citenamefont {Lecordier},\ and\ \citenamefont
  {H\'etet}}]{PhysRevLett.121.053602}%
  \BibitemOpen
  \bibfield  {author} {\bibinfo {author} {\bibfnamefont {T.}~\bibnamefont
  {Delord}}, \bibinfo {author} {\bibfnamefont {P.}~\bibnamefont {Huillery}},
  \bibinfo {author} {\bibfnamefont {L.}~\bibnamefont {Schwab}}, \bibinfo
  {author} {\bibfnamefont {L.}~\bibnamefont {Nicolas}}, \bibinfo {author}
  {\bibfnamefont {L.}~\bibnamefont {Lecordier}}, \ and\ \bibinfo {author}
  {\bibfnamefont {G.}~\bibnamefont {H\'etet}},\ }\href {\doibase
  10.1103/PhysRevLett.121.053602} {\bibfield  {journal} {\bibinfo  {journal}
  {Phys. Rev. Lett.}\ }\textbf {\bibinfo {volume} {121}},\ \bibinfo {pages}
  {053602} (\bibinfo {year} {2018})}\BibitemShut {NoStop}%
\bibitem [{\citenamefont {Conangla}\ \emph {et~al.}(2018)\citenamefont
  {Conangla}, \citenamefont {Schell}, \citenamefont {Rica},\ and\ \citenamefont
  {Quidant}}]{Conangla}%
  \BibitemOpen
  \bibfield  {author} {\bibinfo {author} {\bibfnamefont {G.~P.}\ \bibnamefont
  {Conangla}}, \bibinfo {author} {\bibfnamefont {A.~W.}\ \bibnamefont
  {Schell}}, \bibinfo {author} {\bibfnamefont {R.~A.}\ \bibnamefont {Rica}}, \
  and\ \bibinfo {author} {\bibfnamefont {R.}~\bibnamefont {Quidant}},\
  }\href@noop {} {\bibfield  {journal} {\bibinfo  {journal} {Nano Letters}\ }
  (\bibinfo {year} {2018})}\BibitemShut {NoStop}%
\bibitem [{\citenamefont {{Rabl}}\ \emph {et~al.}(2010)\citenamefont {{Rabl}},
  \citenamefont {{Kolkowitz}}, \citenamefont {{Koppens}}, \citenamefont
  {{Harris}}, \citenamefont {{Zoller}},\ and\ \citenamefont {{Lukin}}}]{Rabl3}%
  \BibitemOpen
  \bibfield  {author} {\bibinfo {author} {\bibfnamefont {P.}~\bibnamefont
  {{Rabl}}}, \bibinfo {author} {\bibfnamefont {S.~J.}\ \bibnamefont
  {{Kolkowitz}}}, \bibinfo {author} {\bibfnamefont {F.~H.~L.}\ \bibnamefont
  {{Koppens}}}, \bibinfo {author} {\bibfnamefont {J.~G.~E.}\ \bibnamefont
  {{Harris}}}, \bibinfo {author} {\bibfnamefont {P.}~\bibnamefont {{Zoller}}},
  \ and\ \bibinfo {author} {\bibfnamefont {M.~D.}\ \bibnamefont {{Lukin}}},\
  }\href {\doibase 10.1038/nphys1679} {\bibfield  {journal} {\bibinfo
  {journal} {Nature Physics}\ }\textbf {\bibinfo {volume} {6}},\ \bibinfo
  {pages} {602} (\bibinfo {year} {2010})}\BibitemShut {NoStop}%
\bibitem [{\citenamefont {Delord}\ \emph {et~al.}(2020)\citenamefont {Delord},
  \citenamefont {Huillery}, \citenamefont {Nicolas},\ and\ \citenamefont
  {H{\'e}tet}}]{DelordNature}%
  \BibitemOpen
  \bibfield  {author} {\bibinfo {author} {\bibfnamefont {T.}~\bibnamefont
  {Delord}}, \bibinfo {author} {\bibfnamefont {P.}~\bibnamefont {Huillery}},
  \bibinfo {author} {\bibfnamefont {L.}~\bibnamefont {Nicolas}}, \ and\
  \bibinfo {author} {\bibfnamefont {G.}~\bibnamefont {H{\'e}tet}},\ }\href@noop
  {} {\bibfield  {journal} {\bibinfo  {journal} {Nature}\ } (\bibinfo {year}
  {2020})}\BibitemShut {NoStop}%
\bibitem [{\citenamefont {Degen}\ \emph {et~al.}(2009)\citenamefont {Degen},
  \citenamefont {Poggio}, \citenamefont {Mamin}, \citenamefont {Rettner},\ and\
  \citenamefont {Rugar}}]{Degen1313}%
  \BibitemOpen
  \bibfield  {author} {\bibinfo {author} {\bibfnamefont {C.~L.}\ \bibnamefont
  {Degen}}, \bibinfo {author} {\bibfnamefont {M.}~\bibnamefont {Poggio}},
  \bibinfo {author} {\bibfnamefont {H.~J.}\ \bibnamefont {Mamin}}, \bibinfo
  {author} {\bibfnamefont {C.~T.}\ \bibnamefont {Rettner}}, \ and\ \bibinfo
  {author} {\bibfnamefont {D.}~\bibnamefont {Rugar}},\ }\href {\doibase
  10.1073/pnas.0812068106} {\bibfield  {journal} {\bibinfo  {journal}
  {Proceedings of the National Academy of Sciences}\ }\textbf {\bibinfo
  {volume} {106}},\ \bibinfo {pages} {1313} (\bibinfo {year} {2009})},\ \Eprint
  {http://arxiv.org/abs/https://www.pnas.org/content/106/5/1313.full.pdf}
  {https://www.pnas.org/content/106/5/1313.full.pdf} \BibitemShut {NoStop}%
\bibitem [{\citenamefont {Rugar}\ \emph {et~al.}(2004)\citenamefont {Rugar},
  \citenamefont {Budakian}, \citenamefont {Mamin},\ and\ \citenamefont
  {Chui}}]{Rugar2004}%
  \BibitemOpen
  \bibfield  {author} {\bibinfo {author} {\bibfnamefont {D.}~\bibnamefont
  {Rugar}}, \bibinfo {author} {\bibfnamefont {R.}~\bibnamefont {Budakian}},
  \bibinfo {author} {\bibfnamefont {H.~J.}\ \bibnamefont {Mamin}}, \ and\
  \bibinfo {author} {\bibfnamefont {B.~W.}\ \bibnamefont {Chui}},\ }\href
  {https://doi.org/10.1038/nature02658} {\bibfield  {journal} {\bibinfo
  {journal} {Nature}\ }\textbf {\bibinfo {volume} {430}},\ \bibinfo {pages}
  {329 EP } (\bibinfo {year} {2004})}\BibitemShut {NoStop}%
\bibitem [{\citenamefont {Rondin}\ \emph {et~al.}(2014)\citenamefont {Rondin},
  \citenamefont {Tetienne}, \citenamefont {Hingant}, \citenamefont {Roch},
  \citenamefont {Maletinsky},\ and\ \citenamefont {Jacques}}]{rondin}%
  \BibitemOpen
  \bibfield  {author} {\bibinfo {author} {\bibfnamefont {L.}~\bibnamefont
  {Rondin}}, \bibinfo {author} {\bibfnamefont {J.-P.}\ \bibnamefont
  {Tetienne}}, \bibinfo {author} {\bibfnamefont {T.}~\bibnamefont {Hingant}},
  \bibinfo {author} {\bibfnamefont {J.-F.}\ \bibnamefont {Roch}}, \bibinfo
  {author} {\bibfnamefont {P.}~\bibnamefont {Maletinsky}}, \ and\ \bibinfo
  {author} {\bibfnamefont {V.}~\bibnamefont {Jacques}},\ }\href@noop {}
  {\bibfield  {journal} {\bibinfo  {journal} {Reports on Progress in Physics}\
  }\textbf {\bibinfo {volume} {77}},\ \bibinfo {pages} {056503} (\bibinfo
  {year} {2014})}\BibitemShut {NoStop}%
\bibitem [{\citenamefont {Robledo}\ \emph {et~al.}(2011)\citenamefont
  {Robledo}, \citenamefont {Childress}, \citenamefont {Bernien}, \citenamefont
  {Hensen}, \citenamefont {Alkemade},\ and\ \citenamefont
  {Hanson}}]{robledo2011high}%
  \BibitemOpen
  \bibfield  {author} {\bibinfo {author} {\bibfnamefont {L.}~\bibnamefont
  {Robledo}}, \bibinfo {author} {\bibfnamefont {L.}~\bibnamefont {Childress}},
  \bibinfo {author} {\bibfnamefont {H.}~\bibnamefont {Bernien}}, \bibinfo
  {author} {\bibfnamefont {B.}~\bibnamefont {Hensen}}, \bibinfo {author}
  {\bibfnamefont {P.~F.}\ \bibnamefont {Alkemade}}, \ and\ \bibinfo {author}
  {\bibfnamefont {R.}~\bibnamefont {Hanson}},\ }\href@noop {} {\bibfield
  {journal} {\bibinfo  {journal} {Nature}\ }\textbf {\bibinfo {volume} {477}},\
  \bibinfo {pages} {574} (\bibinfo {year} {2011})}\BibitemShut {NoStop}%
\bibitem [{\citenamefont {Rao}\ \emph {et~al.}(2016)\citenamefont {Rao},
  \citenamefont {Momenzadeh},\ and\ \citenamefont
  {Wrachtrup}}]{rao2016heralded}%
  \BibitemOpen
  \bibfield  {author} {\bibinfo {author} {\bibfnamefont {D.~B.}\ \bibnamefont
  {Rao}}, \bibinfo {author} {\bibfnamefont {S.~A.}\ \bibnamefont {Momenzadeh}},
  \ and\ \bibinfo {author} {\bibfnamefont {J.}~\bibnamefont {Wrachtrup}},\
  }\href@noop {} {\bibfield  {journal} {\bibinfo  {journal} {Physical review
  letters}\ }\textbf {\bibinfo {volume} {117}},\ \bibinfo {pages} {077203}
  (\bibinfo {year} {2016})}\BibitemShut {NoStop}%
\bibitem [{\citenamefont {Delord}\ \emph
  {et~al.}(2017{\natexlab{c}})\citenamefont {Delord}, \citenamefont {Nicolas},
  \citenamefont {Bodini},\ and\ \citenamefont {H{\'e}tet}}]{vacuumESR}%
  \BibitemOpen
  \bibfield  {author} {\bibinfo {author} {\bibfnamefont {T.}~\bibnamefont
  {Delord}}, \bibinfo {author} {\bibfnamefont {L.}~\bibnamefont {Nicolas}},
  \bibinfo {author} {\bibfnamefont {M.}~\bibnamefont {Bodini}}, \ and\ \bibinfo
  {author} {\bibfnamefont {G.}~\bibnamefont {H{\'e}tet}},\ }\href@noop {}
  {\bibfield  {journal} {\bibinfo  {journal} {Applied Physics Letters}\
  }\textbf {\bibinfo {volume} {111}},\ \bibinfo {pages} {013101} (\bibinfo
  {year} {2017}{\natexlab{c}})}\BibitemShut {NoStop}%
\bibitem [{\citenamefont {Abbott}\ \emph {et~al.}(2007)\citenamefont {Abbott},
  \citenamefont {Ergeneman}, \citenamefont {Kummer}, \citenamefont {Hirt},\
  and\ \citenamefont {Nelson}}]{4392562}%
  \BibitemOpen
  \bibfield  {author} {\bibinfo {author} {\bibfnamefont {J.~J.}\ \bibnamefont
  {Abbott}}, \bibinfo {author} {\bibfnamefont {O.}~\bibnamefont {Ergeneman}},
  \bibinfo {author} {\bibfnamefont {M.~P.}\ \bibnamefont {Kummer}}, \bibinfo
  {author} {\bibfnamefont {A.~M.}\ \bibnamefont {Hirt}}, \ and\ \bibinfo
  {author} {\bibfnamefont {B.~J.}\ \bibnamefont {Nelson}},\ }\href {\doibase
  10.1109/TRO.2007.910775} {\bibfield  {journal} {\bibinfo  {journal} {IEEE
  Transactions on Robotics}\ }\textbf {\bibinfo {volume} {23}},\ \bibinfo
  {pages} {1247} (\bibinfo {year} {2007})}\BibitemShut {NoStop}%
\bibitem [{\citenamefont {Osborn}(1945)}]{PhysRev.67.351}%
  \BibitemOpen
  \bibfield  {author} {\bibinfo {author} {\bibfnamefont {J.~A.}\ \bibnamefont
  {Osborn}},\ }\href {\doibase 10.1103/PhysRev.67.351} {\bibfield  {journal}
  {\bibinfo  {journal} {Phys. Rev.}\ }\textbf {\bibinfo {volume} {67}},\
  \bibinfo {pages} {351} (\bibinfo {year} {1945})}\BibitemShut {NoStop}%
\bibitem [{\citenamefont {Shearwood}\ \emph {et~al.}(2000)\citenamefont
  {Shearwood}, \citenamefont {Ho}, \citenamefont {Williams},\ and\
  \citenamefont {Gong}}]{Shearwood}%
  \BibitemOpen
  \bibfield  {author} {\bibinfo {author} {\bibfnamefont {C.}~\bibnamefont
  {Shearwood}}, \bibinfo {author} {\bibfnamefont {K.~Y.}\ \bibnamefont {Ho}},
  \bibinfo {author} {\bibfnamefont {C.~B.}\ \bibnamefont {Williams}}, \ and\
  \bibinfo {author} {\bibfnamefont {H.}~\bibnamefont {Gong}},\ }\href {\doibase
  https://doi.org/10.1016/S0924-4247(00)00292-2} {\bibfield  {journal}
  {\bibinfo  {journal} {Sensors and Actuators A: Physical}\ }\textbf {\bibinfo
  {volume} {83}},\ \bibinfo {pages} {85} (\bibinfo {year} {2000})}\BibitemShut
  {NoStop}%
\bibitem [{\citenamefont {Wu}\ \emph {et~al.}(2014)\citenamefont {Wu},
  \citenamefont {Hryciw}, \citenamefont {Healey}, \citenamefont {Lake},
  \citenamefont {Jayakumar}, \citenamefont {Freeman}, \citenamefont {Davis},\
  and\ \citenamefont {Barclay}}]{PhysRevX.4.021052}%
  \BibitemOpen
  \bibfield  {author} {\bibinfo {author} {\bibfnamefont {M.}~\bibnamefont
  {Wu}}, \bibinfo {author} {\bibfnamefont {A.~C.}\ \bibnamefont {Hryciw}},
  \bibinfo {author} {\bibfnamefont {C.}~\bibnamefont {Healey}}, \bibinfo
  {author} {\bibfnamefont {D.~P.}\ \bibnamefont {Lake}}, \bibinfo {author}
  {\bibfnamefont {H.}~\bibnamefont {Jayakumar}}, \bibinfo {author}
  {\bibfnamefont {M.~R.}\ \bibnamefont {Freeman}}, \bibinfo {author}
  {\bibfnamefont {J.~P.}\ \bibnamefont {Davis}}, \ and\ \bibinfo {author}
  {\bibfnamefont {P.~E.}\ \bibnamefont {Barclay}},\ }\href {\doibase
  10.1103/PhysRevX.4.021052} {\bibfield  {journal} {\bibinfo  {journal} {Phys.
  Rev. X}\ }\textbf {\bibinfo {volume} {4}},\ \bibinfo {pages} {021052}
  (\bibinfo {year} {2014})}\BibitemShut {NoStop}%
\bibitem [{\citenamefont {Kim}\ \emph {et~al.}(2016{\natexlab{a}})\citenamefont
  {Kim}, \citenamefont {Hauer}, \citenamefont {Doolin}, \citenamefont
  {Souris},\ and\ \citenamefont {Davis}}]{Kim2016}%
  \BibitemOpen
  \bibfield  {author} {\bibinfo {author} {\bibfnamefont {P.~H.}\ \bibnamefont
  {Kim}}, \bibinfo {author} {\bibfnamefont {B.~D.}\ \bibnamefont {Hauer}},
  \bibinfo {author} {\bibfnamefont {C.}~\bibnamefont {Doolin}}, \bibinfo
  {author} {\bibfnamefont {F.}~\bibnamefont {Souris}}, \ and\ \bibinfo {author}
  {\bibfnamefont {J.~P.}\ \bibnamefont {Davis}},\ }\href
  {https://doi.org/10.1038/ncomms13165} {\bibfield  {journal} {\bibinfo
  {journal} {Nature Communications}\ }\textbf {\bibinfo {volume} {7}},\
  \bibinfo {pages} {13165 EP } (\bibinfo {year} {2016}{\natexlab{a}})},\
  \bibinfo {note} {article}\BibitemShut {NoStop}%
\bibitem [{\citenamefont {Jain}\ \emph {et~al.}(2016)\citenamefont {Jain},
  \citenamefont {Gieseler}, \citenamefont {Moritz}, \citenamefont {Dellago},
  \citenamefont {Quidant},\ and\ \citenamefont {Novotny}}]{jain2016direct}%
  \BibitemOpen
  \bibfield  {author} {\bibinfo {author} {\bibfnamefont {V.}~\bibnamefont
  {Jain}}, \bibinfo {author} {\bibfnamefont {J.}~\bibnamefont {Gieseler}},
  \bibinfo {author} {\bibfnamefont {C.}~\bibnamefont {Moritz}}, \bibinfo
  {author} {\bibfnamefont {C.}~\bibnamefont {Dellago}}, \bibinfo {author}
  {\bibfnamefont {R.}~\bibnamefont {Quidant}}, \ and\ \bibinfo {author}
  {\bibfnamefont {L.}~\bibnamefont {Novotny}},\ }\href@noop {} {\bibfield
  {journal} {\bibinfo  {journal} {Physical review letters}\ }\textbf {\bibinfo
  {volume} {116}},\ \bibinfo {pages} {243601} (\bibinfo {year}
  {2016})}\BibitemShut {NoStop}%
\bibitem [{\citenamefont {Wang}\ \emph {et~al.}(2019)\citenamefont {Wang},
  \citenamefont {Lourette}, \citenamefont {O'Kelley}, \citenamefont {Kayci},
  \citenamefont {Band}, \citenamefont {Kimball}, \citenamefont {Sushkov},\ and\
  \citenamefont {Budker}}]{Wang19}%
  \BibitemOpen
  \bibfield  {author} {\bibinfo {author} {\bibfnamefont {T.}~\bibnamefont
  {Wang}}, \bibinfo {author} {\bibfnamefont {S.}~\bibnamefont {Lourette}},
  \bibinfo {author} {\bibfnamefont {S.~R.}\ \bibnamefont {O'Kelley}}, \bibinfo
  {author} {\bibfnamefont {M.}~\bibnamefont {Kayci}}, \bibinfo {author}
  {\bibfnamefont {Y.}~\bibnamefont {Band}}, \bibinfo {author} {\bibfnamefont
  {D.~F.~J.}\ \bibnamefont {Kimball}}, \bibinfo {author} {\bibfnamefont
  {A.~O.}\ \bibnamefont {Sushkov}}, \ and\ \bibinfo {author} {\bibfnamefont
  {D.}~\bibnamefont {Budker}},\ }\href {\doibase
  10.1103/PhysRevApplied.11.044041} {\bibfield  {journal} {\bibinfo  {journal}
  {Phys. Rev. Applied}\ }\textbf {\bibinfo {volume} {11}},\ \bibinfo {pages}
  {044041} (\bibinfo {year} {2019})}\BibitemShut {NoStop}%
\bibitem [{\citenamefont {{Rusconi}}\ \emph {et~al.}(2017)\citenamefont
  {{Rusconi}}, \citenamefont {{P{\"o}chhacker}}, \citenamefont {{Kustura}},
  \citenamefont {{Cirac}},\ and\ \citenamefont
  {{Romero-Isart}}}]{2017PhRvL.119p7202R}%
  \BibitemOpen
  \bibfield  {author} {\bibinfo {author} {\bibfnamefont {C.~C.}\ \bibnamefont
  {{Rusconi}}}, \bibinfo {author} {\bibfnamefont {V.}~\bibnamefont
  {{P{\"o}chhacker}}}, \bibinfo {author} {\bibfnamefont {K.}~\bibnamefont
  {{Kustura}}}, \bibinfo {author} {\bibfnamefont {J.~I.}\ \bibnamefont
  {{Cirac}}}, \ and\ \bibinfo {author} {\bibfnamefont {O.}~\bibnamefont
  {{Romero-Isart}}},\ }\href {\doibase 10.1103/PhysRevLett.119.167202}
  {\bibfield  {journal} {\bibinfo  {journal} {\prl}\ }\textbf {\bibinfo
  {volume} {119}},\ \bibinfo {eid} {167202} (\bibinfo {year} {2017})},\ \Eprint
  {http://arxiv.org/abs/1703.09346} {arXiv:1703.09346 [quant-ph]} \BibitemShut
  {NoStop}%
\bibitem [{\citenamefont {Jackson~Kimball}\ \emph {et~al.}(2016)\citenamefont
  {Jackson~Kimball}, \citenamefont {Sushkov},\ and\ \citenamefont
  {Budker}}]{Kimball}%
  \BibitemOpen
  \bibfield  {author} {\bibinfo {author} {\bibfnamefont {D.~F.}\ \bibnamefont
  {Jackson~Kimball}}, \bibinfo {author} {\bibfnamefont {A.~O.}\ \bibnamefont
  {Sushkov}}, \ and\ \bibinfo {author} {\bibfnamefont {D.}~\bibnamefont
  {Budker}},\ }\href {\doibase 10.1103/PhysRevLett.116.190801} {\bibfield
  {journal} {\bibinfo  {journal} {Phys. Rev. Lett.}\ }\textbf {\bibinfo
  {volume} {116}},\ \bibinfo {pages} {190801} (\bibinfo {year}
  {2016})}\BibitemShut {NoStop}%
\bibitem [{\citenamefont {Ohno}\ \emph {et~al.}(2012)\citenamefont {Ohno},
  \citenamefont {Joseph~Heremans}, \citenamefont {Bassett}, \citenamefont
  {Myers}, \citenamefont {Toyli}, \citenamefont {Bleszynski~Jayich},
  \citenamefont {Palmstr{\o}m},\ and\ \citenamefont {Awschalom}}]{Ohno}%
  \BibitemOpen
  \bibfield  {author} {\bibinfo {author} {\bibfnamefont {K.}~\bibnamefont
  {Ohno}}, \bibinfo {author} {\bibfnamefont {F.}~\bibnamefont
  {Joseph~Heremans}}, \bibinfo {author} {\bibfnamefont {L.~C.}\ \bibnamefont
  {Bassett}}, \bibinfo {author} {\bibfnamefont {B.~A.}\ \bibnamefont {Myers}},
  \bibinfo {author} {\bibfnamefont {D.~M.}\ \bibnamefont {Toyli}}, \bibinfo
  {author} {\bibfnamefont {A.~C.}\ \bibnamefont {Bleszynski~Jayich}}, \bibinfo
  {author} {\bibfnamefont {C.~J.}\ \bibnamefont {Palmstr{\o}m}}, \ and\
  \bibinfo {author} {\bibfnamefont {D.~D.}\ \bibnamefont {Awschalom}},\
  }\href@noop {} {\bibfield  {journal} {\bibinfo  {journal} {Applied Physics
  Letters}\ }\textbf {\bibinfo {volume} {101}},\ \bibinfo {pages} {082413}
  (\bibinfo {year} {2012})}\BibitemShut {NoStop}%
\bibitem [{\citenamefont {Tetienne}\ \emph {et~al.}(2012)\citenamefont
  {Tetienne}, \citenamefont {Rondin}, \citenamefont {Spinicelli}, \citenamefont
  {Chipaux}, \citenamefont {Debuisschert}, \citenamefont {Roch},\ and\
  \citenamefont {Jacques}}]{Tetienne2}%
  \BibitemOpen
  \bibfield  {author} {\bibinfo {author} {\bibfnamefont {J.-P.}\ \bibnamefont
  {Tetienne}}, \bibinfo {author} {\bibfnamefont {L.}~\bibnamefont {Rondin}},
  \bibinfo {author} {\bibfnamefont {P.}~\bibnamefont {Spinicelli}}, \bibinfo
  {author} {\bibfnamefont {M.}~\bibnamefont {Chipaux}}, \bibinfo {author}
  {\bibfnamefont {T.}~\bibnamefont {Debuisschert}}, \bibinfo {author}
  {\bibfnamefont {J.-F.}\ \bibnamefont {Roch}}, \ and\ \bibinfo {author}
  {\bibfnamefont {V.}~\bibnamefont {Jacques}},\ }\href@noop {} {\bibfield
  {journal} {\bibinfo  {journal} {New Journal of Physics}\ }\textbf {\bibinfo
  {volume} {14}},\ \bibinfo {pages} {103033} (\bibinfo {year}
  {2012})}\BibitemShut {NoStop}%
\bibitem [{\citenamefont {Law}\ and\ \citenamefont
  {Eberly}(1996{\natexlab{a}})}]{law1996arbitraryb}%
  \BibitemOpen
  \bibfield  {author} {\bibinfo {author} {\bibfnamefont {C.}~\bibnamefont
  {Law}}\ and\ \bibinfo {author} {\bibfnamefont {J.}~\bibnamefont {Eberly}},\
  }\href@noop {} {\bibfield  {journal} {\bibinfo  {journal} {Physical review
  letters}\ }\textbf {\bibinfo {volume} {76}},\ \bibinfo {pages} {1055}
  (\bibinfo {year} {1996}{\natexlab{a}})}\BibitemShut {NoStop}%
\bibitem [{\citenamefont {Hensen}\ \emph {et~al.}(2015)\citenamefont {Hensen},
  \citenamefont {Bernien}, \citenamefont {Dr{\'e}au}, \citenamefont {Reiserer},
  \citenamefont {Kalb}, \citenamefont {Blok}, \citenamefont {Ruitenberg},
  \citenamefont {Vermeulen}, \citenamefont {Schouten}, \citenamefont
  {Abell{\'a}n} \emph {et~al.}}]{hensen2015loophole}%
  \BibitemOpen
  \bibfield  {author} {\bibinfo {author} {\bibfnamefont {B.}~\bibnamefont
  {Hensen}}, \bibinfo {author} {\bibfnamefont {H.}~\bibnamefont {Bernien}},
  \bibinfo {author} {\bibfnamefont {A.~E.}\ \bibnamefont {Dr{\'e}au}}, \bibinfo
  {author} {\bibfnamefont {A.}~\bibnamefont {Reiserer}}, \bibinfo {author}
  {\bibfnamefont {N.}~\bibnamefont {Kalb}}, \bibinfo {author} {\bibfnamefont
  {M.~S.}\ \bibnamefont {Blok}}, \bibinfo {author} {\bibfnamefont
  {J.}~\bibnamefont {Ruitenberg}}, \bibinfo {author} {\bibfnamefont {R.~F.}\
  \bibnamefont {Vermeulen}}, \bibinfo {author} {\bibfnamefont {R.~N.}\
  \bibnamefont {Schouten}}, \bibinfo {author} {\bibfnamefont {C.}~\bibnamefont
  {Abell{\'a}n}},  \emph {et~al.},\ }\href@noop {} {\bibfield  {journal}
  {\bibinfo  {journal} {Nature}\ }\textbf {\bibinfo {volume} {526}},\ \bibinfo
  {pages} {682} (\bibinfo {year} {2015})}\BibitemShut {NoStop}%
\bibitem [{\citenamefont {Jarmola}\ \emph {et~al.}(2012)\citenamefont
  {Jarmola}, \citenamefont {Acosta}, \citenamefont {Jensen}, \citenamefont
  {Chemerisov},\ and\ \citenamefont {Budker}}]{Jarmola}%
  \BibitemOpen
  \bibfield  {author} {\bibinfo {author} {\bibfnamefont {A.}~\bibnamefont
  {Jarmola}}, \bibinfo {author} {\bibfnamefont {V.~M.}\ \bibnamefont {Acosta}},
  \bibinfo {author} {\bibfnamefont {K.}~\bibnamefont {Jensen}}, \bibinfo
  {author} {\bibfnamefont {S.}~\bibnamefont {Chemerisov}}, \ and\ \bibinfo
  {author} {\bibfnamefont {D.}~\bibnamefont {Budker}},\ }\href@noop {}
  {\bibfield  {journal} {\bibinfo  {journal} {Phys. Rev. Lett.}\ }\textbf
  {\bibinfo {volume} {108}},\ \bibinfo {pages} {197601} (\bibinfo {year}
  {2012})}\BibitemShut {NoStop}%
\bibitem [{\citenamefont {Law}\ and\ \citenamefont
  {Eberly}(1996{\natexlab{b}})}]{LawEberly}%
  \BibitemOpen
  \bibfield  {author} {\bibinfo {author} {\bibfnamefont {C.~K.}\ \bibnamefont
  {Law}}\ and\ \bibinfo {author} {\bibfnamefont {J.~H.}\ \bibnamefont
  {Eberly}},\ }\href {\doibase 10.1103/PhysRevLett.76.1055} {\bibfield
  {journal} {\bibinfo  {journal} {Phys. Rev. Lett.}\ }\textbf {\bibinfo
  {volume} {76}},\ \bibinfo {pages} {1055} (\bibinfo {year}
  {1996}{\natexlab{b}})}\BibitemShut {NoStop}%
\bibitem [{\citenamefont {Stickler}\ \emph {et~al.}(2016)\citenamefont
  {Stickler}, \citenamefont {Papendell},\ and\ \citenamefont
  {Hornberger}}]{stickler2016spatio}%
  \BibitemOpen
  \bibfield  {author} {\bibinfo {author} {\bibfnamefont {B.~A.}\ \bibnamefont
  {Stickler}}, \bibinfo {author} {\bibfnamefont {B.}~\bibnamefont {Papendell}},
  \ and\ \bibinfo {author} {\bibfnamefont {K.}~\bibnamefont {Hornberger}},\
  }\href@noop {} {\bibfield  {journal} {\bibinfo  {journal} {Physical Review
  A}\ }\textbf {\bibinfo {volume} {94}},\ \bibinfo {pages} {033828} (\bibinfo
  {year} {2016})}\BibitemShut {NoStop}%
\bibitem [{\citenamefont {Zhong}\ and\ \citenamefont
  {Robicheaux}(2016)}]{zhong2016decoherence}%
  \BibitemOpen
  \bibfield  {author} {\bibinfo {author} {\bibfnamefont {C.}~\bibnamefont
  {Zhong}}\ and\ \bibinfo {author} {\bibfnamefont {F.}~\bibnamefont
  {Robicheaux}},\ }\href@noop {} {\bibfield  {journal} {\bibinfo  {journal}
  {Physical Review A}\ }\textbf {\bibinfo {volume} {94}},\ \bibinfo {pages}
  {052109} (\bibinfo {year} {2016})}\BibitemShut {NoStop}%
\bibitem [{\citenamefont {Fremerey}(1982)}]{fremerey1982spinning}%
  \BibitemOpen
  \bibfield  {author} {\bibinfo {author} {\bibfnamefont {J.}~\bibnamefont
  {Fremerey}},\ }\href@noop {} {\bibfield  {journal} {\bibinfo  {journal}
  {Vacuum}\ }\textbf {\bibinfo {volume} {32}},\ \bibinfo {pages} {685}
  (\bibinfo {year} {1982})}\BibitemShut {NoStop}%
\bibitem [{\citenamefont {Mizuochi}\ \emph {et~al.}(2009)\citenamefont
  {Mizuochi}, \citenamefont {Neumann}, \citenamefont {Rempp}, \citenamefont
  {Beck}, \citenamefont {Jacques}, \citenamefont {Siyushev}, \citenamefont
  {Nakamura}, \citenamefont {Twitchen}, \citenamefont {Watanabe}, \citenamefont
  {Yamasaki} \emph {et~al.}}]{mizuochi2009coherence}%
  \BibitemOpen
  \bibfield  {author} {\bibinfo {author} {\bibfnamefont {N.}~\bibnamefont
  {Mizuochi}}, \bibinfo {author} {\bibfnamefont {P.}~\bibnamefont {Neumann}},
  \bibinfo {author} {\bibfnamefont {F.}~\bibnamefont {Rempp}}, \bibinfo
  {author} {\bibfnamefont {J.}~\bibnamefont {Beck}}, \bibinfo {author}
  {\bibfnamefont {V.}~\bibnamefont {Jacques}}, \bibinfo {author} {\bibfnamefont
  {P.}~\bibnamefont {Siyushev}}, \bibinfo {author} {\bibfnamefont
  {K.}~\bibnamefont {Nakamura}}, \bibinfo {author} {\bibfnamefont
  {D.}~\bibnamefont {Twitchen}}, \bibinfo {author} {\bibfnamefont
  {H.}~\bibnamefont {Watanabe}}, \bibinfo {author} {\bibfnamefont
  {S.}~\bibnamefont {Yamasaki}},  \emph {et~al.},\ }\href@noop {} {\bibfield
  {journal} {\bibinfo  {journal} {Physical review B}\ }\textbf {\bibinfo
  {volume} {80}},\ \bibinfo {pages} {041201} (\bibinfo {year}
  {2009})}\BibitemShut {NoStop}%
\bibitem [{\citenamefont {Rosskopf}\ \emph {et~al.}(2014)\citenamefont
  {Rosskopf}, \citenamefont {Dussaux}, \citenamefont {Ohashi}, \citenamefont
  {Loretz}, \citenamefont {Schirhagl}, \citenamefont {Watanabe}, \citenamefont
  {Shikata}, \citenamefont {Itoh},\ and\ \citenamefont {Degen}}]{Rosskopf}%
  \BibitemOpen
  \bibfield  {author} {\bibinfo {author} {\bibfnamefont {T.}~\bibnamefont
  {Rosskopf}}, \bibinfo {author} {\bibfnamefont {A.}~\bibnamefont {Dussaux}},
  \bibinfo {author} {\bibfnamefont {K.}~\bibnamefont {Ohashi}}, \bibinfo
  {author} {\bibfnamefont {M.}~\bibnamefont {Loretz}}, \bibinfo {author}
  {\bibfnamefont {R.}~\bibnamefont {Schirhagl}}, \bibinfo {author}
  {\bibfnamefont {H.}~\bibnamefont {Watanabe}}, \bibinfo {author}
  {\bibfnamefont {S.}~\bibnamefont {Shikata}}, \bibinfo {author} {\bibfnamefont
  {K.~M.}\ \bibnamefont {Itoh}}, \ and\ \bibinfo {author} {\bibfnamefont
  {C.~L.}\ \bibnamefont {Degen}},\ }\href {\doibase
  10.1103/PhysRevLett.112.147602} {\bibfield  {journal} {\bibinfo  {journal}
  {Phys. Rev. Lett.}\ }\textbf {\bibinfo {volume} {112}},\ \bibinfo {pages}
  {147602} (\bibinfo {year} {2014})}\BibitemShut {NoStop}%
\bibitem [{\citenamefont {Ostroff}\ and\ \citenamefont
  {Waugh}(1966)}]{Ostroff}%
  \BibitemOpen
  \bibfield  {author} {\bibinfo {author} {\bibfnamefont {E.~D.}\ \bibnamefont
  {Ostroff}}\ and\ \bibinfo {author} {\bibfnamefont {J.~S.}\ \bibnamefont
  {Waugh}},\ }\href {\doibase 10.1103/PhysRevLett.16.1097} {\bibfield
  {journal} {\bibinfo  {journal} {Phys. Rev. Lett.}\ }\textbf {\bibinfo
  {volume} {16}},\ \bibinfo {pages} {1097} (\bibinfo {year}
  {1966})}\BibitemShut {NoStop}%
\bibitem [{\citenamefont {Bluvstein}\ \emph {et~al.}(2019)\citenamefont
  {Bluvstein}, \citenamefont {Zhang}, \citenamefont {McLellan}, \citenamefont
  {Williams},\ and\ \citenamefont {Jayich}}]{bluvstein}%
  \BibitemOpen
  \bibfield  {author} {\bibinfo {author} {\bibfnamefont {D.}~\bibnamefont
  {Bluvstein}}, \bibinfo {author} {\bibfnamefont {Z.}~\bibnamefont {Zhang}},
  \bibinfo {author} {\bibfnamefont {C.~A.}\ \bibnamefont {McLellan}}, \bibinfo
  {author} {\bibfnamefont {N.~R.}\ \bibnamefont {Williams}}, \ and\ \bibinfo
  {author} {\bibfnamefont {A.~C.~B.}\ \bibnamefont {Jayich}},\ }\href {\doibase
  10.1103/PhysRevLett.123.146804} {\bibfield  {journal} {\bibinfo  {journal}
  {Phys. Rev. Lett.}\ }\textbf {\bibinfo {volume} {123}},\ \bibinfo {pages}
  {146804} (\bibinfo {year} {2019})}\BibitemShut {NoStop}%
\bibitem [{\citenamefont {Kim}\ \emph {et~al.}(2016{\natexlab{b}})\citenamefont
  {Kim}, \citenamefont {Hauer}, \citenamefont {Doolin}, \citenamefont
  {Souris},\ and\ \citenamefont {Davis}}]{Kim}%
  \BibitemOpen
  \bibfield  {author} {\bibinfo {author} {\bibfnamefont {P.~H.}\ \bibnamefont
  {Kim}}, \bibinfo {author} {\bibfnamefont {B.~D.}\ \bibnamefont {Hauer}},
  \bibinfo {author} {\bibfnamefont {C.}~\bibnamefont {Doolin}}, \bibinfo
  {author} {\bibfnamefont {F.}~\bibnamefont {Souris}}, \ and\ \bibinfo {author}
  {\bibfnamefont {J.~P.}\ \bibnamefont {Davis}},\ }\href@noop {} {\bibfield
  {journal} {\bibinfo  {journal} {Nature Communications}\ }\textbf {\bibinfo
  {volume} {7}},\ \bibinfo {pages} {13165 EP } (\bibinfo {year}
  {2016}{\natexlab{b}})}\BibitemShut {NoStop}%
\bibitem [{\citenamefont {Chen}\ and\ \citenamefont
  {Yin}(2019{\natexlab{b}})}]{Chen}%
  \BibitemOpen
  \bibfield  {author} {\bibinfo {author} {\bibfnamefont {X.-Y.}\ \bibnamefont
  {Chen}}\ and\ \bibinfo {author} {\bibfnamefont {Z.-q.}\ \bibnamefont {Yin}},\
  }\href@noop {} {\bibfield  {journal} {\bibinfo  {journal} {Phys. Rev. A}\
  }\textbf {\bibinfo {volume} {99}},\ \bibinfo {pages} {022319} (\bibinfo
  {year} {2019}{\natexlab{b}})}\BibitemShut {NoStop}%
\end{thebibliography}
\end{document}